\documentclass[twocolumn]{aastex631}

\newcommand{\gs}{M_{\rm HI}/M_*}

\shorttitle{Gas in BreakBRDs}
\shortauthors{Stark et al.}
\graphicspath{{./}{figures/}}

\begin{document}

\title{BreakBRD Galaxies: Evolutionary Clues Through an Analysis of Gas Content} 

\author[0000-0002-0786-7307]{David V. Stark}
\affiliation{Space Telescope Science Institute \\ 3700 San Martin Dr \\ Baltimore, MD 21218, USA}
\affiliation{William H. Miller III Department of Physics and Astronomy\\ Johns Hopkins University\\ Baltimore, MD 21218, USA}

\author{Sarah Tuttle}
\affiliation{Department of Astronomy, University of Washington \\
3910 15th Ave. NE, Room C319 \\
Seattle, WA 98195-0002, USA}

\author{Stephanie Tonnesen}
\affiliation{CCA, Flatiron Institute\\ 162 5th Avenue\\ New York, NY 10010, USA}

\author{Zachary Tu}
\affiliation{Department of Astronomy, University of Washington \\
3910 15th Ave. NE, Room C319 \\
Seattle, WA 98195-0002, USA}

\author{Sean P. Fillingham}
\affiliation{Department of Astronomy, University of Washington \\
3910 15th Ave. NE, Room C319 \\
Seattle, WA 98195-0002, USA}



\begin{abstract}
By combining newly obtained deep GBT 21cm observations with optical spectroscopic data, we present an analysis of the gas content of BreakBRD galaxies, a population denoted by their blue star-forming centers and red quenched disks that do not appear to follow the typical inside-out evolution of spiral galaxies. We confirm previous results that the neutral atomic hydrogen (\ion{H}{1}) gas fractions of BreakBRDs are on-average lower than those of typical galaxies on the star-forming sequence (SFS), and find that their \ion{H}{1} fractions are generally higher than Green Valley (GV) galaxies. \ion{H}{1} depletion times for BreakBRDs are roughly an order of magnitude lower than those of SFS galaxies, in stark contrast with GV galaxies that typically have much longer depletion times than SFS galaxies. The nuclear gas-phase metallicities of BreakBRDs have a broader distribution than SFS galaxies and are skewed towards slightly above-average values. BreakBRDs are systematically offset from the Baryonic Tully-Fisher Relation towards lower baryonic mass at a given rotation velocity. They also have higher typical \ion{H}{1} asymmetries than SFS galaxies, and of those galaxies with spatially resolved gas velocity fields from the SDSS-IV MaNGA survey, two-thirds are either highly distorted or completely misaligned relative to the stellar disk. Evidence supports a scenario where BreakBRDs are in an early phase of quenching, and there is mixed evidence that their behavior is related to past merger activity. 
\end{abstract}

\keywords{}


\section{Introduction} \label{sec:intro}

 One of the most fundamental ways galaxies can be categorized is through their star-formation activity. Since the peak of cosmic star formation around $z\sim2$, the non star-forming (quenched) population, also referred to as the ``red sequence" based on its position in color-magnitude space, has grown significantly \citep{Bell2004, Faber2007}. While numerous mechanisms have been identified or proposed to drive quenching, large statistical studies of galaxy populations have shown that quenching is a function of both stellar mass and environment \citep{Peng2010}, and that it typically occurs ``inside-out," where galaxy bulges tend to quench before disks \citep{White91, Chiappini1997,VandenBosch1998,Boissier1999,Perez2013, LopezFernandez2018,Li2015,ErrozFerrer2019,Lin2019}.     

However, the inside-out mechanism is not the only way quenching can manifest. Outside-in quenching has also been observed, and such behavior is generally expected from environmental processes like ram-pressure stripping that preferentially impact disks \citep{Rasmussen2006,Koopmann2004a, Koopmann2004b}. Environmental processes can affect all galaxies, but they are expected to more severely impact low-mass galaxies with shallower potential wells \citep[e.g., ][]{Kolcu2022}. However, observational evidence for this behavior is inconsistent; \citet{Schaefer2017} find a link between outside-in quenching and dense environments, while \citet{Welikala2008} find local environments tend to drive the reverse. \citet{Spindler2018} see environmentally-driven quenching independent of galactocentric radius. Alternatively, tidal interactions and bar instabilities can cause gas to be redistributed from the disk to the central regions of galaxies \citep{Masters10, Ellison11, Moreno2015}, which could lead to red quenched disks and blue star forming bulges, at least temporarily, if enough material were to flow inwards. The ability of mergers and/or bars to drive gas inwards may be mass-dependent, being more impactful for massive galaxies \citep{Ellison11, Carles2016}. At the same time, mergers are also thought to be responsible for the buildup of spheroidal components, which may drive inside-out quenching by stabilizing any gas in the central regions \citep{Martig2009}. Inflows will also potentially drive active galactic nuclei (AGN) activity that will likely preferentially affect central regions as well.

One specific class of galaxies that shows apparent outside-in evolution is the ``Break Bulges in Red Disks" (BreakBRD) sample introduced by \citet{Tuttle20}. This subsample is defined as having recent central star formation identified via the 4000$\AA$ break (hence, ``break bulges") but optically red disks inconsistent with recent star formation. \citet{Tuttle20} found that this population is ubiquitous in both stellar mass and environment parameter space, indicating that no single mass- or environment-dependent process is driving their behavior, and that their gas content is lower than that of typical star-forming galaxies. Radial star formation profiles derived from a subset of BreakBRDs with SDSS-IV MaNGA \citep{Bundy15} integral field unit (IFU) spectroscopic data  has confirmed that their disks are indeed older than their bulges and their bulges are still actively star-forming, but that star formation is not enhanced above what is expected from the star-forming sequence (McKay et al., in preparation). 

\citet{Kopenhafer2020} identified BreakBRD analogs in the IllustrisTNG cosmological hydrodynamic simulation, finding a higher rate of satellite and splashback galaxies among BreakBRDs, but that there nonetheless exists a significant fraction of BreakBRDs that are central in their halos and have no clear driver for their behavior. They also found lower (but not zero) and more centrally concentrated star formation, a more concentrated distribution of dense star-forming gas, and a history of gas loss in their outskirts since $z\sim0.5$.  Most of their BreakBRD analogs were quenching systems. \cite{Tonnesen2023} followed this work with an analysis of the simulated CGM, finding it to be less massive, more compact, and more metal rich than typical star-forming galaxies. 

These past studies support a scenario where BreakBRDs are quenching from the outside-in, possibly due to a gas disk that has either been lost or stabilized against star formation.   The original analysis of BreakBRDs in \citet{Tuttle20} only had a limited, shallow census of the gas reservoir, namely the neutral atomic hydrogen (\ion{H}{1}). Although \ion{H}{1} is not the direct fuel for active star-formation, it is the large-scale reservoir out of which star-forming molecular clouds condense, making it an important quantity in order to understand the long-term star formation potential of a galaxy. Furthermore, as \ion{H}{1} is often extended relative to stellar disks, it can be a sensitive tracer of external influences \citep{Holwerda11}, making it useful for determining what may be driving BreakBRD evolution. In this study, we present new 21cm observations for a sample of BreakBRDs. These data provide robust measurements with either detections or strong upper limits (\ion{H}{1}-to-stellar mass ratio $\lesssim$ 0.05) over most of the sample.  Our new 21cm observations are combined with optical spectroscopic data from the SDSS Legacy Spectroscopic survey \citep{York00} and the SDSS-IV MaNGA survey for a subset of fourteen BreakBRDs. With these data, we analyze the \ion{H}{1} properties, nuclear gas-phase elemental abundances, and spatially resolved gas kinematics of breakBRDs, all in an effort to more clearly constrain how they fit into the galaxy evolution paradigm.

This paper is organized as follows. In Section~\ref{sec:data} we discuss the BreakBRD and control samples comprised of typical star-forming and quenching galaxies, the \ion{H}{1} data, and optical spectroscopic data. In Section~\ref{sec:results} we present our results, specifically examining \ion{H}{1} fractions and depletion times, nuclear gas-phase metallicities, and gas kinematics of BreakBRDs relative to our control samples. In Section~\ref{sec:discussion}, we discuss how our results inform whether BreakBRDs are indeed a unique subset of quenching galaxies, and what may be driving their behavior. A summary of our results is presented in Section~\ref{sec:conclusions}. Throughout this paper we assume a flat cosmology with $H_0=70\,{\rm km\,s^{-1}\,kpc^{-1}}$, $\Omega_m=0.3$, and $\Omega_{\Lambda}=0.7$.

\section{Sample and Data} \label{sec:data}
\subsection{The BreakBRD sample} \label{sec:bbrd_def}
BreakBRD galaxies are characterized as having star-forming bulges and quiescent disks. They are defined as having no ongoing AGN activity. For our study, we start with the same sample of BreakBRDs presented in \citet{Tuttle20}, whose selection we briefly summarize here. The BreakBRDs are selected from a sample of $\sim$70,000 local ($z<$0.05) face-on SDSS galaxies used for bulge/disk decomposition by \citet{Lackner2012}. To ensure robust bulge-disk decomposition, the parent sample is first limited to galaxies with absolute $r$-band magnitude $M_r<-17$, axial ratio $b/a > 0.7$, and bulge/disk colors of $0.2<g-r<0.9$. The small fraction (125 total) of galaxies with non star-forming disks ($g-r>0.6$) and recent central star formation (based on $D_N 4000$) are the BreakBRDs. The sample is further refined by rejecting galaxies classified as AGN using [\ion{O}{3}]/H$\beta$ vs. [\ion{N}{2}]/H$\alpha$ Balwin, Phillips, and Tellervich (BPT) diagrams \citep{Baldwin81, Kewley06b}. 

\subsection{HI data} \label{sec:data_hi}
Global \ion{H}{1} properties used in this study are derived from 21cm observations carried out with single-dish radio telescopes. A total of 73 BreakBRD galaxies fall within the footprint of the Arecibo Legacy Fast ALFA (ALFALFA) survey \citep{Haynes18}. These cases were identified by crossmatching the BreakBRD optical positions with the ALFALFA \ion{H}{1} positions from the $\alpha$.100 source catalog using a match radius of 2{\arcmin}. For any match on-sky, we also require the optical redshift to fall within 1.5 times the \ion{H}{1} $W_{50}$ linewidth from the \ion{H}{1} redshift. For galaxies falling within the ALFALFA footprint but lacking counterparts in the ALFALFA catalog, we extract each of their spectra from the original ALFALFA data cubes in $4 \arcmin \times 4 \arcmin$ boxes centered on each galaxy position. We then measure the $rms$ noise directly in the spectra and estimate an \ion{H}{1} upper limit using Equation~4 of \citet{Stark21} assuming an intrinsic galaxy linewidth of $200\,{\rm km\,s^{-1}}$.

ALFALFA observations had a fixed integration time of less than one minute per beam, which, combined with the required S/N of 4.5 for detections in the $\alpha$.100 catalog, yielded limited sensitivity to \ion{H}{1} mass at the distance of the breakBRD sample. Only 29\% of the 73 breakBRDs had published ALFALFA measurements. For galaxies lacking ALFALFA data, or those that had weak ALFALFA detections ($S/N<5$) or upper limits ($M_{\rm HI}/M_* > 0.05$), were observed using the Robert C. Byrd Green Bank Telescope (GBT) under project codes AGBT20B-261 and AGBT22A-436. In total, 98 galaxies were observed. The L-band receiver and Versatile GBT Astronomical Spectrometer (VEGAS) backend were used. Observations were performed using position switching with five minute scans. Integration times varied, set either by the time needed to reach an integrated flux $S/N > 5$ or an \ion{H}{1} upper limit of $M_{\rm HI}/M_* < 0.05$. Our approach resulted in 80\% of breakBRDs observed with the GBT yielding detections. Periodic observations of bright radio standards were used to check the pointing and flux calibration of our observations (see Appendix~A for further information).

GBT data were reduced using the pipeline designed for the HI-MaNGA survey \citep{Masters19,Stark21}, which we briefly summarize here. Each scan is inspected for contamination from Global Positioning Satellite (GPS) interference at 1385 MHz (a recession velocity of $\sim8500\,{\rm km\,s^{-1}}$ for the \ion{H}{1} rest frequency). Any of the ten second sub-integrations containing this signal are rejected. Scans are then averaged and any remaining narrow radio frequency interference (RFI) features flagged. The data are boxcar and hanning smoothed to a spectral resolution of $\sim10\,{\rm km\,s^{-1}}$. A polynomial baseline is fit and subtracted from each spectrum using only signal-free regions. The rms noise is measured from a large contiguous, flat, and signal-free region of the baseline-subtracted spectrum. For spectra with clear 21cm signals (based on a by-eye inspection of the spectrum at the expected location of the 21cm emission line), the integrated flux, systemic velocity, and linewidth are measured. For non-detections, a flux upper limit is estimated assuming a linewidth of $200\,{\rm km\,s^{-1}}$. 

Cases of likely source confusion are also identified following the methodology outlined in \citet{Stark21}. Cases where multiple optical counterparts lie within a 1.5 beam radius of each observation while also having optical redshifts within 1.5$W_{50}$ of the detected \ion{H}{1} emission line velocity centroid are flagged as possibly confused. In a second step, the \ion{H}{1} mass likelihood distribution is estimated for each optical counterpart using $u-i$ color, stellar mass, and mean $r$-band surface brightness, which is then converted into an observed 21cm flux likelihood using the galaxies' luminosity distances and angular separation from the beam center (the latter accounts for the declining beam sensitivity away from its center). The likelihood distributions are then used to estimate the probability that at least 20\% of the \ion{H}{1} flux in an observation comes from galaxies other than the primary target. Any galaxies where this probability is $>$10\% are removed from our analysis. 37/98 (38\%) GBT observations and 8/73 (11\%) ALFALFA observations are flagged as confused via this step.

A sample of our complete catalog containing both ALFALFA and GBT measurements is presented in Table~\ref{tbl:catalog}. The full version is available as a machine readable table in the online version of this article. It contains 98 GBT observations and 73 ALFALFA observations for 118 unique galaxies. Users should be aware that many galaxies have entries for both GBT and ALFALFA data; even when the GBT data is deeper, we include the parameters from the ALFALFA survey for completeness and because use-cases for the catalog may differ between users. From our data set, we generate a final catalog for our analysis (with one measurement per galaxy) using the following decision tree to determine which observation to use in cases where a galaxy has both GBT and ALFALFA data: (1) If both observations are {\it unconfused} detections, the one with the higher $S/N$ is used. (2) If both observations are non-detections, the one yielding the stronger upper limit is used. (3) If one observation is a confused detection and the other an unconfused detection, the unconfused detection is used. (4) If one observation is an {\it unconfused} detection and the other an upper limit, the detection is used. (5) If one observation is a {\it confused} detection and the other an upper limit, the upper limit is typically used, except for a handful of cases where the flux measured from the confused detection is actually lower than the upper limit (this can occur in cases with very deep GBT data and relatively shallow ALFALFA data), in which case the confused detection is adopted as a flux upper limit. (6) If both observations are confused detections, neither is used in any analysis of \ion{H}{1} properties.

The following is a complete description of the columns in the catalog. 
\begin{itemize}
    \item SDSS ID: The identifier for a given galaxy.
    \item $V_{\rm opt}$: optical heliocentric recession velocity.
    \item $t_{on}$: Total on-source time for a given galaxy. This does not include the time spent in the OFF position.
    \item $\sigma_{rms}$: rms noise measured in the spectrum after boxcar and hanning smoothing to a resolution of ${\sim}10\,{\rm km\,s^{-1}}$. 
    \item $F_{\rm HI}$: total integrated 21cm flux.
    \item $\sigma_{F_{HI}}$: uncertainty on integrated 21cm flux.
    \item $f_{\rm peak}$: peak flux density of the 21cm profile.
    \item $\log{M_{\rm HI}}$: Integrated \ion{H}{1} mass for detections calculated using Equation 5 from \citet{Stark21}.
    \item $\log{M_{\rm HI, lim}}$: Derived \ion{H}{1} upper limit for non-detections assuming an intrinsic linewidth of $200\,{\rm km\,s^{-1}}$, calculated using Equations 3 or 4 from \citet{Stark21} for GBT and ALFALFA data, respectively.
    \item $W_{F50}$: \ion{H}{1} linewidth measured between the points where the flux density is equal to half of the peak flux density on each side. The peak flux densities on each side of the profile are typically measured at the nearest ``horn" for a classic two-horned profile, or at the central peak for a Gaussian-shaped profile. The channel where the flux density equals half the peak value is estimated by fitting a line on either side of the profile for the data with flux density between 0.1-0.9 times the peak flux density. The linewidths are corrected for instrumental broadening following \citet{Springob05}, but are not corrected for turbulence.
    \item $V_{\rm HI}$: Heliocentric recession velocity derived from the \ion{H}{1} emission line. This is calculated by taking the midpoint of where the flux density reaches half of its peak value on either side of the profile (where the peak flux density is either from the nearest ``horn" for a two-horned profile, or the central peak for a Gaussian-shaped profile).
    \item $p_{conf}$: Probability that the \ion{H}{1} data are confused estimated following the methodology described in \citet{Stark21}. The value specifically conveys the likelihood that at least 20\% of the \ion{H}{1} flux arises from a galaxy other than the primary target. 
    \item source: Source of the 21cm observation. G: Green Bank Telescope; A: ALFALFA survey (Arecibo telescope; \citealt{Haynes18}).
\end{itemize}

\begin{rotatetable*}
\movetabledown=5cm
\begin{deluxetable*}{ccccccccccccccc} \label{tbl:catalog}
\tablecaption{\ion{H}{1} Measurements for the BreakBRD sample$^*$}
\tablehead{SDSS ID & $V_{\rm opt}$ & $t_{\rm on}$ & $\sigma_{rms}$ & $F_{\rm HI}$ & $\sigma_{F_{\rm HI}}$ & $f_{\rm peak}$ & $M_{\rm HI}$ & $ M_{\rm HI, lim}$ & $v_{\rm HI}$ & $W_{\rm F50}$ & $p_{\rm conf}$ & source\tablenotemark{a} \\
 & $\rm km\,s^{-1}$, & s & mJy & $\rm Jy\,km\,s^{-1}$ & $\rm Jy\,km\,s^{-1}$ & mJy & $\log{M_{\odot}}$ & $\log{M_{\odot}}$ & ${\rm km\,s^{-1}}$ & ${\rm km\,s^{-1}}$}
\startdata
SDSS J011344.94$+$154100.3 & 13191 & 1195 & 1.35 & 0.74 & 0.09 & 5.92 & 9.79 & \nodata & 13284 & 207 & 0.93 & G \\
SDSS J011728.10$+$144215.9 & 12891 & 2989 & 0.71 & 0.84 & 0.06 & 4.41 & 9.81 & \nodata & 12823 & 292 & 0.00 & G \\
SDSS J011858.59-002606.4 & 14090 & 5976 & 0.57 & 0.17 & 0.03 & 2.07 & 9.21 & \nodata & 14227 & 104 & 0.00 & G \\
SDSS J012132.95-005811.7 & 14690 & 6576 & 0.64 & 0.32 & 0.04 & 2.78 & 9.51 & \nodata & 14611 & 224 & 0.00 & G \\
SDSS J074602.40$+$184301.7 & 13848 & 597 & 1.88 & 1.76 & 0.17 & 11.39 & 10.20 & \nodata & 13863 & 369 & 0.89 & G \\
SDSS J075008.55$+$500243.1 & 7041 & 557 & 1.68 & 1.21 & 0.12 & 8.65 & 9.45 & \nodata & 7035 & 211 & 0.00 & G \\
SDSS J075158.32$+$154241.0 & 14090 & 10465 & 0.48 & \nodata & \nodata & \nodata & \nodata & 8.95 & \nodata & \nodata & 0.00 & G \\
SDSS J075822.53$+$121208.8 & 14990 & 897 & 1.74 & 2.82 & 0.18 & 9.31 & 10.47 & \nodata & 14876 & 434 & 0.96 & G \\
SDSS J080643.29$+$453938.0 & 10193 & 9049 & 0.49 & \nodata & \nodata & \nodata & \nodata & 8.68 & \nodata & \nodata & 0.00 & G \\
SDSS J082127.82$+$122333.3 & 12591 & 5372 & 0.75 & 0.24 & 0.05 & 1.97 & 9.25 & \nodata & 12492 & 235 & 0.00 & G \\
\enddata
\tablenotetext{a}{G: GBT, A: ALFALFA}
\tablenotetext{*}{This table is available in its entirety in the online version of this manuscript.}
\end{deluxetable*}
\end{rotatetable*}

\subsection{Optical Spectroscopic Data} \label{sec:data_manga}

Fourteen BreakBRD galaxies have optical integral field unit (IFU) spectroscopic data from the Sloan Digital Sky Survey (SDSS-IV) MaNGA Nearby Galaxies at Apache Point Observatory (MaNGA) survey \citep{Bundy15}. We use the final release (SDSS DR17) data products generated by the latest MaNGA Data Analysis Pipeline (DAP, \citealt{Westfall19}). Our study makes use of the stellar and gas kinematic maps generated by this pipeline.

For our full breakBRD sample (not just the fourteen with MaNGA data), we also use spectroscopic measurements from the SDSS Legacy Survey \citep{York00}, specifically making use of several emission line fluxes. These data are limited to the central 3{\arcsec} of each galaxy so do not necessarily reflect global galaxy conditions.

\subsection{Global Star Formation Rates and Stellar Masses}
For consistency, total galaxy star formation rates and stellar masses are taken from \citet{Chang15}, who model spectral energy distributions (SEDs) for the full SDSS Legacy spectroscopic sample using SDSS optical and WISE infrared photometry. The star formation rates from this analysis importantly incorporate dust-obscured star formation. We use the median of the SFR and $M_*$ likelihood distributions generated for each galaxy. 

\subsection{Control Sample Generation}
Control samples for comparison to the BreakBRD sample are drawn from the SDSS-IV MaNGA survey, specifically the subset of MaNGA with global 21cm measurements from the HI-MaNGA follow-up survey \citep{Masters19, Stark21}. Two basic control samples are generated. The first is composed of ``typical" star-forming galaxies that lie on the ``Star Forming (Main) Sequence", selected as having EW(H$\alpha$)$>6\AA$ measured within their half-light radii using MaNGA IFU data \citep{Sanchez14,Cano-Diaz19} and [\ion{N}{2}]/H$\alpha$ and [\ion{O}{3}]/H$\beta$ line ratios inconsistent with AGN. This control sample is given the shorthand, ``SFS" throughout this paper. The second control sample is composed of galaxies in the ``Green Valley" between star forming and quenched galaxies, selected as having $3<$EW(H$\alpha$)$<6\AA$, and is given the shorthand, ``GV" throughout this paper. As has been discussed in \citet{Tuttle20}, and further in our own study, BreakBRDs contain star formation, but show evidence of possible quenching. The star-forming and green-valley control samples provide a means of understanding how BreakBRDs are consistent with, or distinct from, these two key stages of galaxy evolution.

For each analysis, we customize the control sample generation depending on which galaxy parameters must be fixed. In some cases, it may be sufficient to simply match on stellar mass, while in other cases, we may need to match other properties (e.g., redshift, axial ratio) to avoid bias caused by secondary observational effects. In all cases, we always match in stellar mass to within 0.2 dex of each primary galaxy, with the value of 0.2 dex chosen since it is a typical systematic stellar mass uncertainty \citep{Kannappan07}. If any additional properties are fixed for a given analysis, they will be mentioned in the corresponding text in Section~\ref{sec:results}.

To improve the statistical strength of our analysis, when possible we generate control samples that are larger than the primary sample but have the same distribution of properties. To generate these control samples, for each primary galaxy, we draw a unique random control galaxy from all available galaxies with properties similar to the primary. This process is repeated until there are no remaining control galaxies for {\it all} galaxies in the primary sample. In the figures demonstrating comparisons between the breakBRD and control samples, the number of galaxies in each sample is provided in the figure caption.

\section{Results} \label{sec:results}
In the following section we present the results of our analysis investigating how the gas content and properties of BreakBRDs differ from the population of SFS and GV galaxies. In Section~\ref{sec:results_gas}, we explore differences in global HI-to-stellar mass ratios and \ion{H}{1} depletion times. In Section~\ref{sec:results_metal}, we investigate central gas-phase metallicities, and in Section~\ref{sec:results_vel}, we identify key differences in gas kinematic properties. 

\subsection{\ion{H}{1} Fractions and Depletion Times} \label{sec:results_gas}

\begin{figure*}
\includegraphics[width=0.5\textwidth]{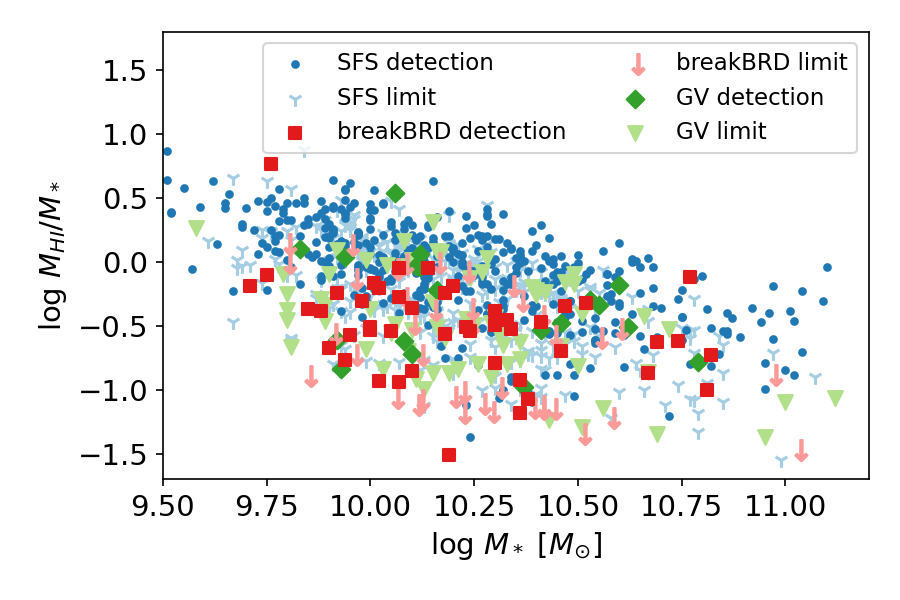}
\includegraphics[width=0.5\textwidth]{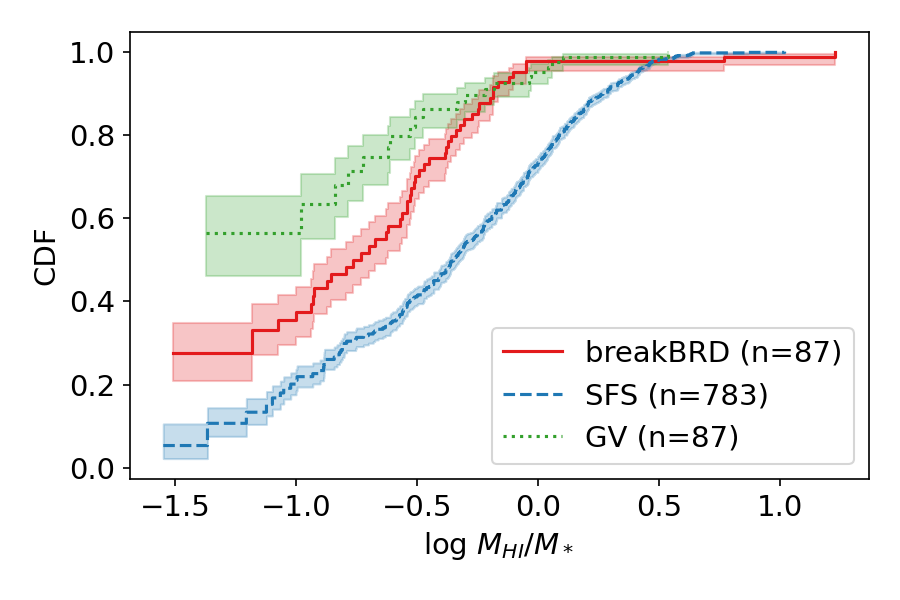}
\caption{(left) HI-to-stellar mass ratio versus stellar mass for BreakBRD galaxies and the SFS and GV control samples. \ion{H}{1} detections and non-detections are shown for each sample. (right) The cumulative distribution function (CDF) of HI-to-stellar mass ratio for the BreakBRDs and the control samples, determined using the Kaplan-Meier estimator. Shaded regions indicate the 68\% confidence intervals. BreakBRDs have typcial \ion{H}{1}-to-stellar mass ratios intermediate between SFS and GV galaxies.} 
\label{fig:gs_mstars}
\end{figure*}

The left panel of Figure~\ref{fig:gs_mstars} shows the \ion{H}{1}-to-stellar mass ratio ($\gs$) versus stellar mass ($M_*$) relationship for all BreakBRDs and the corresponding control samples, which are matched in stellar mass. The total numbers in each sample are indicated in the right-hand panel of Figure~\ref{fig:gs_mstars}. All the plotted samples contain significant numbers of upper limits, limiting the comparison of the $\gs$ distributions to some degree, but it is clear that the BreakBRD sample does not generally extend to as high $\gs$ as the SFS control sample, and it appears to have a distribution roughly comparable to the GV sample.  To further investigate the distributions, the right panel of Figure~\ref{fig:gs_mstars} plots the cumulative distribution function (CDF) of $\gs$ for the three samples. Due to the presence of upper limits, we employ survival analysis methods to estimate the CDF, specifically using the Kaplan-Meier estimator from the \texttt{lifelines} Python package \citep{Davidson-Pilon2019} to determine the survival function, $S$, for $\gs$, from which the CDF is calculated as $1-S$. It can be clearly seen that the CDF for BreakBRD galaxies has a median value of $\gs$ roughly 0.5 dex below that of the SFS control sample. This analysis confirms earlier results from \citet{Tuttle20} that used a smaller subset of data from just ALFALFA, but the deeper \ion{H}{1} observations also show that BreakBRDs are not uniformly gas-poor either. At fixed stellar mass, their gas fractions still span $\sim$1 dex. The CDF for the GV control does not have a well-constrained mean value due to the large fraction of upper limits, but the portion of the CDF we do constrain implies lower typical gas fractions compared to BreakBRDs. Therefore, BreakBRDs appear to have typical gas fractions intermediate between those in the Green Valley and the Star Forming Sequence.

We next investigate gas consumption timescales in BreakBRDs using the \ion{H}{1} depletion time, $\tau_{\rm HI}$, defined as
\begin{equation}
    \tau_{HI} = M_{HI}/SFR
\end{equation}
where $SFR$ is the global star formation rate. Although this metric makes simplistic assumptions (e.g, constant SFR, no gas replenishment, and all gas available for star formation) it provides a rough indicator of the future star formation potential of a galaxy.

\begin{figure*}
\includegraphics[width=0.5\textwidth]{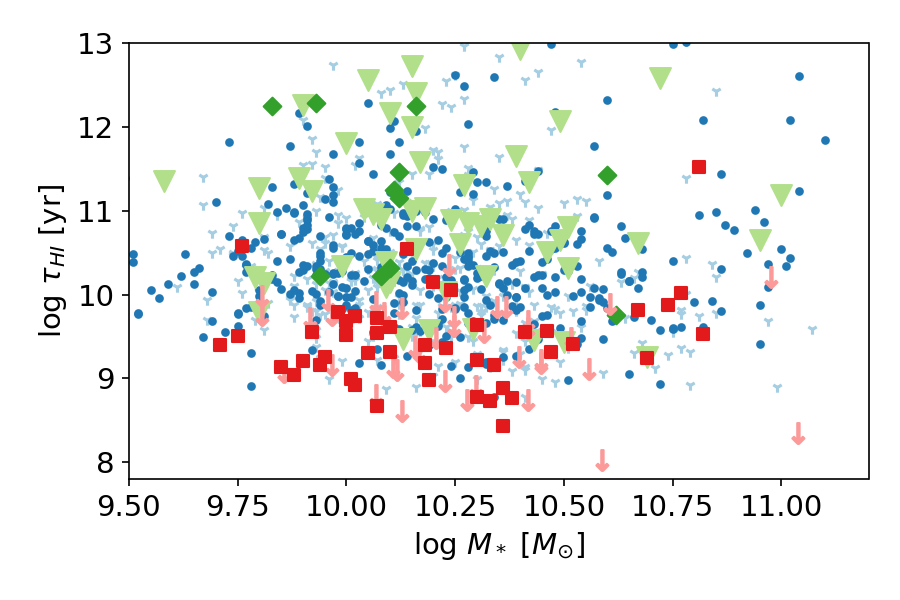}
\includegraphics[width=0.5\textwidth]{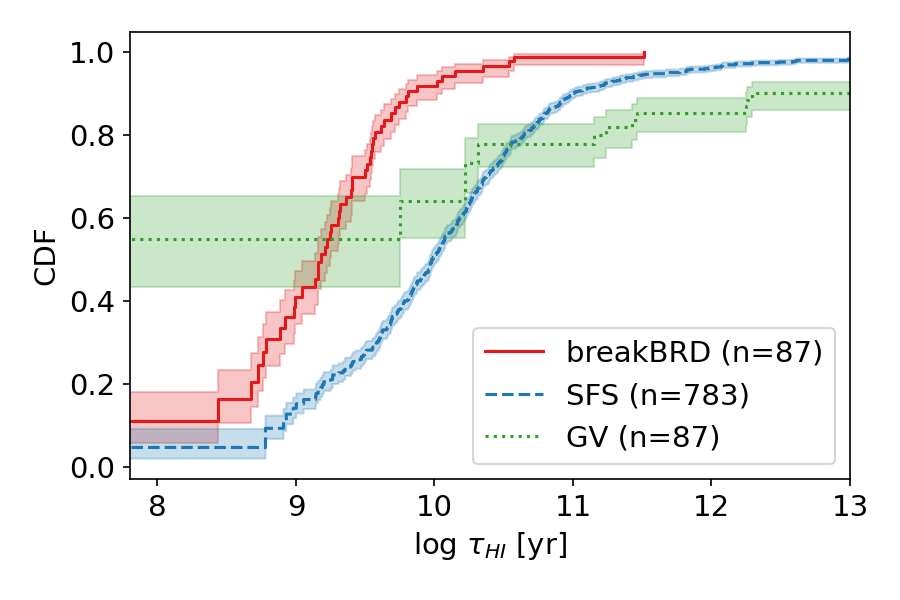}
\caption{(left) \ion{H}{1} depletion time versus stellar mass for BreakBRD galaxies and the SFS and GV control samples. Detections and non-detections are shown for each sample. Symbols are the same as in Figure~\ref{fig:gs_mstars}. (right) CDFs of \ion{H}{1} depletion time for the three samples, again determined using the Kaplan-Meier estimator. Shaded regions indicate the 68\% confidence interval. BreakBRDs have \ion{H}{1} depletion times that are shorter than both SFS and GV galaxies.} 
\label{fig:tdep_mstars}
\end{figure*}

The left panel of Figure~\ref{fig:tdep_mstars} shows $\tau_{HI}$ vs. stellar mass for BreakBRDs as well as the SFS and GV control samples, which are identical to those used to study the $\gs$ vs. $M_*$ relation. As in Figure~\ref{fig:gs_mstars} there are large numbers of upper limits, but the distribution of $\tau_{HI}$ for the SFS control sample clearly extends to higher values generally unpopulated by BreakBRDs. The right panel of this figure shows the estimated CDFs, which emphasizes this point further. Using these CDFs, we estimate a median $\tau_{HI}$ of just over 1 Gyr for BreakBRDs, whereas the SFS control sample has a median $\tau_{HI}\sim10\,{\rm Gyr}$.

The distribution of $\tau_{HI}$ for the GV control sample is very distinct from that of the BreakBRDs. Although the details of the GV sample \ion{H}{1} distribution carry large uncertainties, it is still clear that the GV population contains a larger fraction of galaxies with very {\it long} depletion times ($>>10$ Gyr) compared to the SFS and especially the BreakBRD samples. The long depletion times of GV galaxies occur despite them having the lowest typical \ion{H}{1} fraction, implying that the long depletion times are the result of extremely low SFRs. This is distinct from BreakBRDs which have \ion{H}{1} fractions that are lower but global SFRs that are broadly consistent with SFS galaxies, implying it is the decrease in \ion{H}{1} content, not SFR, that is driving their {\it short} depletion times. The implications of this interpretation will be discussed further in Section~\ref{sec:discussion}.

\subsection{Gas Phase Metallicity} \label{sec:results_metal}

We next examine the gas-phase metallicities of BreakBRDs, specifically focusing on \textit{central} metallicities obtained via optical spectroscopic measurements in the central 3$\arcsec$ of all galaxies from the SDSS Legacy spectroscopic survey.

Nuclear gas phase metallicities ($12+\log{O/H}$) are estimated using the O3N2 strong line method of \citet{Marino13} that combines the [\ion{O}{3}]5008\AA/H$\beta$ and [\ion{N}{2}]6585\AA/H$\alpha$ line ratios. To ensure the metallicity calibration is being applied appropriately, we exclude any galaxies whose line ratios fall on the LI(N)ER or Seyfert region of the [\ion{O}{3}]/H$\beta$ vs. [\ion{N}{2}]/H$\alpha$ and [\ion{O}{3}]/H$\beta$ vs. [\ion{S}{2}]/H$\alpha$ BPT diagrams, yielding a total of 77 BreakBRD galaxies for this analysis. The majority of galaxies excluded here are classified as ``composite". Despite cutting the sample, Kolmogorov-Smirnov (K-S) tests performed on the stellar mass and SFR distributions indicate that the subset of BreakBRDs used here is a representative subset of the full BreakBRD sample. Unlike in Section~\ref{sec:results_gas}, we do not require \ion{H}{1} observations for this analysis. The control samples are matched in stellar mass. Due to the combination of weaker emission lines and larger fractions of AGN/Seyfert/LINER BPT classifications among the GV sample, there are fewer galaxies eligible for our analysis. A total of 29 BreakBRDs had no viable match among the GV sample. Therefore, we separate our analysis of BreakBRD and GV galaxies from that of BreakBRDs and SFS galaxies, only plotting the BreakBRDs with available matches in the former.

\begin{figure*}
\plottwo{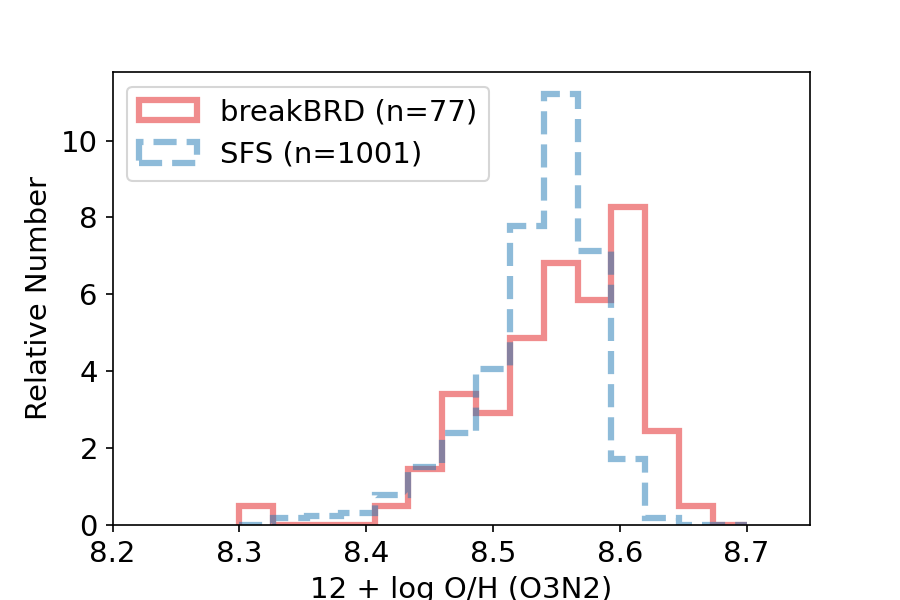}{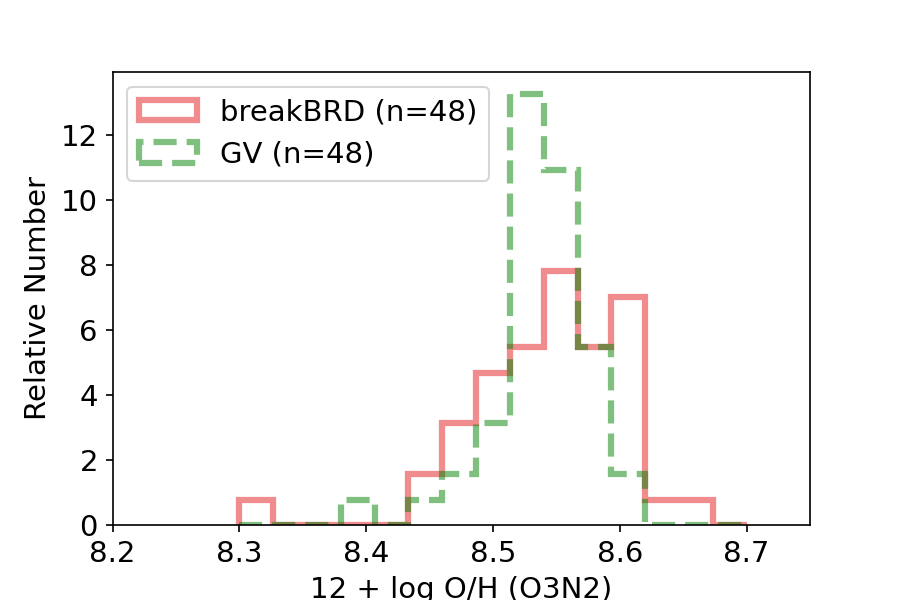}
\caption{Normalized distribution of central SDSS fiber metallicity measurements for BreakBRDs alongside the SFS control (left) and GV control (right) samples. BreakBRDs have a broader distribution of gas-phase metallicity compared to SFS galaxies, and the distribution is skewed towards above-average values. BreakBRDs and GV galaxies appear to have consistent metallicity distributions.} 
\label{fig:central_metal}
\end{figure*}

Figure~\ref{fig:central_metal} shows the distribution of nuclear gas-phase metallicity for BreakBRDs and the control samples. While both the BreakBRD and SFS control sample distributions have approximately equal mean metallicity of $12+\log{O/H}\sim8.55$, the BreakBRD metallicity distribution is significantly more skewed towards larger metallicities (Anderson-Darling and Shapiro-Wilk tests confirm the nuclear metallicity distribution deviates significantly from Normal). A K-S test on the nuclear metallicity distribution of the two samples returns only a $5\times10^{-5}$\% chance that they follow the same distribution. Meanwhile, the BreakBRD and GV samples again have very similar mean values, and a K-S test returns a p-value of 0.25, indicates their metallicity distributions are the same.

\subsection{Gas and Stellar Kinematics} \label{sec:results_vel}
Integrated 21cm observations not only provide information regarding total gas content, but also basic kinematic information through linewidths and spectral profile asymmetries, although it should be noted that any spectral profile asymmetry could be due to asymmetries in kinematics {\it or} gas density distributions. Spatially resolved velocity fields using optical IFU data are available for a small subset of our sample, and provide an even more powerful approach to identifying disturbances in the internal motions of galaxies. This information can provide valuable clues as to what is driving the evolution of galaxies.

\subsubsection{The Baryonic Tully-Fisher Relation} 
\label{sec:btfr}

The Baryonic Tully-Fisher Relation (BTFR) is a fundamental scaling relation between baryonic mass (in this case, stellar plus cold gas mass) and rotation velocity for disk galaxies. The BTFR holds over several orders of magnitude, from gas-rich dwarfs to giant galaxies \citep{McGaugh00, Stark2009}.  Therefore, deviations may highlight galaxies undergoing unique evolutionary processes. 

\begin{figure*}
\plotone{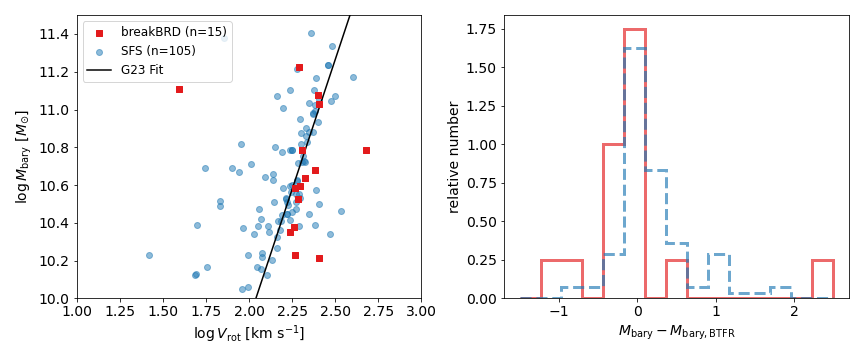}
\caption{(left) The BTFR for BreakBRDs and a control sample of SFS galaxies. GV galaxies are not analyzed due the low rate of strong \ion{H}{1} detections needed for reliable $W_{F50}$ measurements. The fitted relation is from \citep{Goddy2023}. (right) Baryonic mass residuals from the BTFR for the two samples. BreakBRDs fall systematically below the BTFR.}
\label{fig:w50}
\end{figure*}

In Figure~\ref{fig:w50}, we plot the BTFR for 15 BreakBRDs and a control sample of 105 SFS galaxies matched on stellar mass and axial ratio. Matching on axial ratio is to ensure that differences in the distribution of inclination do not lead to any spurious results, although both samples are inclination-corrected. The small sample size used here is largely due to the requirement of peak $S/N>5$ on \ion{H}{1} detections (where peak $S/N = f_{\rm peak}/\sigma_{\rm rms}$), which was enforced because linewidths become significantly more unreliable at lower $S/N$ \citep{Stark13}. We also require the detections to not be confused. We do not plot the GV control sample here as there are too few galaxies with strong enough \ion{H}{1} detections. 

We plot the BTFR following the methodology of \citet{Goddy2023}. Rotation velocity is calculated as $V_{\rm rot}=W_{F50}/2\sin{i}$ where inclination is estimated using Equation 5 from \citet{Masters19}. To match the baryonic mass used in \citet{Goddy2023}, we add 0.24 dex to the logarithm of our stellar masses to convert from a Chabrier to Salpeter IMF, and we multiply our stellar masses by a factor of 1.07 as an approximation for the contribution of molecular hydrogen to the baryonic mass budget. Lastly, we multiply our \ion{H}{1} masses by 4/3 to correct for the contribution of Helium. In Figure~\ref{fig:w50}, we overplot the derived forward fitted BTFR from \citep{Goddy2023} that does not apply a turbulence correction to the velocity measurements (to be consistent with our own data), and is reproduced here for convenience:
\begin{equation}
    M_{\rm bary} = 2.75 V_{\rm rot}+4.39
\end{equation}
The forward ordinary least-squares fit is purposefully used instead of the bisector fit because we will examine baryonic mass residuals from the BTFR (the forward fit minimizes scatter along this axis).

Residuals from the BTFR are shown in the right panel of Figure~\ref{fig:w50}. A K-S test applied to the residuals for BreakBRDs and the SFS control samples indicates they are significantly different with only a 1.2\% chance they follow the same underlying distributions. We find median residuals of $-0.14 \pm 0.07$ and $0.08 \pm 0.03$ for BreakBRDs and SFS galaxies, respectively, and the standard deviations of the residuals are estimated to be $0.72 \pm 0.26$ and $0.57 \pm 0.09$. Uncertainties on these quantities are estimated from bootstrapping. We conclude that BreakBRDs sytematically deviate from typical star forming galaxies around the BTFR, having less total baryonic mass for their rotation velocity. 

\subsubsection{\ion{H}{1} Profile Asymmetry} \label{sec:hi_asym}

We next examine the distribution of 21cm profile asymmetries in our samples. Asymmetries can indicate a lopsidedness in either the \ion{H}{1} distribution or its kinematics, and may be indicative of a number of processes including tidal interactions, gas accretion, and/or disk instabilities (see Section~\ref{sec:discussion_dynamics} for futher discussion). In this work, \ion{H}{1} profile asymmetry, $A_{HI}$, is defined as the logarithmic ratio of 21cm flux above and below a galaxy's systemic velocity:
\begin{equation}
    A = \log{\frac{F_l}{F_h}}
\end{equation}
where $F_l$ and $F_h$ are the integrated fluxes above and below the systemic velocity. The use of a logarithm is mostly for convenience when calculating this quantity; it ensures the absolute magnitude of the asymmetry is independent of whether $F_l$ or $F_h$ is larger.

The choice of how to define the systemic velocity of the \ion{H}{1} profile is an important one. Most past studies of integrated \ion{H}{1} asymmetry use the systemic velocity derived from the \ion{H}{1} profile itself. This may be, for example, a luminosity weighted mean over the spectral profile or the midpoint between the two sides of the profile where the flux density reaches 50\% of the maximum value. Regardless, defining a galaxy's systemic velocity from the \ion{H}{1} profile itself runs the risk of hiding the real asymmetry because the \ion{H}{1} systemic velocity is typically defined to be near the center of the \ion{H}{1} distribution. Alternatively, one may use an independent measure of a galaxy's systemic velocity, and in our case, SDSS fiber spectroscopy provides a useful alternative, although this approach is not without its own set of potential systematic errors. For example, while these measurements are nominally conducted on the photometric centers of galaxies, the optical center may not exactly match the kinematic center of a given galaxy. We opt to conduct our analysis both ways, first defining the systemic velocity from the \ion{H}{1} profile ($V_{HI}$), and second, using the independent redshift measured from optical SDSS single fiber spectroscopy. The former runs the risk of underestimating the true \ion{H}{1} asymmetry, while the latter runs the risk of overestimating the true \ion{H}{1} asymmetry, so they likely both bracket the true values, and if trends are seen using both definitions, they are more likely robust. The first definition of \ion{H}{1} asymmetry also enables better comparison to past work, while we see the latter as less biased. 

For our analysis, we again limit ourselves to cases with \ion{H}{1} peak $S/N > 5$ to ensure reliable estimation of the systemic \ion{H}{1} velocity. The SFS control sample is matched in stellar mass and axial ratio, the latter because \ion{H}{1} profile shape, and the potential of seeing \ion{H}{1} asymmetries, is inclination-dependent. Once again, we can only consider non-confused detections here. We do not analyze GV galaxies here due to the lack of high $S/N$ \ion{H}{1} detections.

\begin{figure*}
    \plottwo{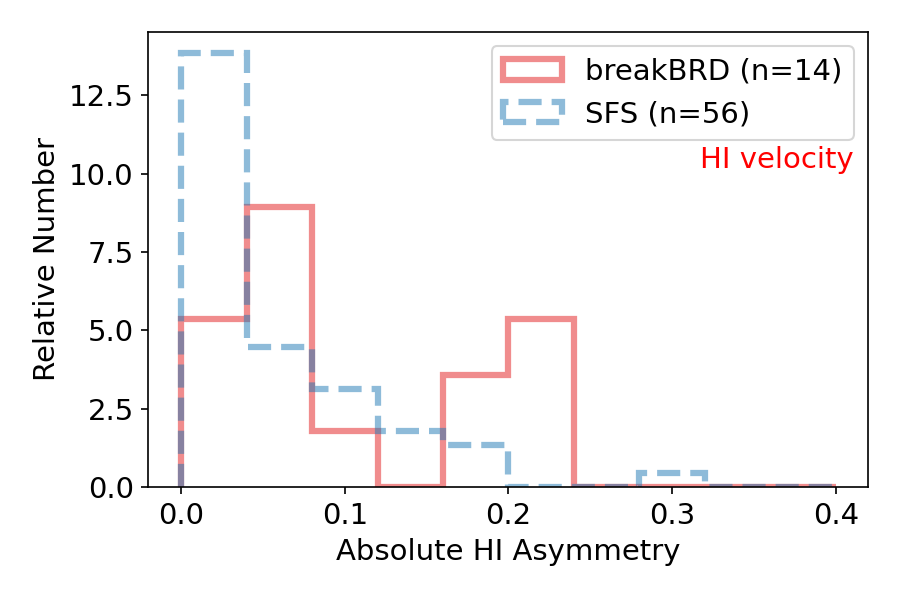}{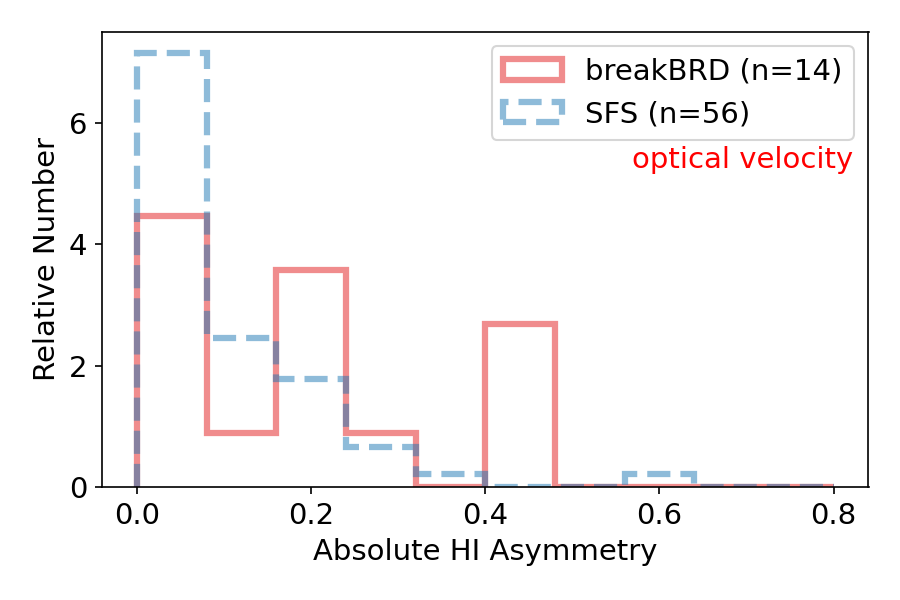}
    \caption{Distribution of \ion{H}{1} profile asymmetries for BreakBRD and SFS samples, defining the systemetic velocity using the \ion{H}{1} profile itself (left) and velocity estimated from optical spectroscopy (right). Both cases show more high-asymmetry profiles among the BreakBRD galaxies than the SFS galaxies.}
    \label{fig:hi_asym}
\end{figure*}

Figure~\ref{fig:hi_asym} shows the distribution of the absolute magnitude of asymmetry for the BreakBRD and SFS sample under both asymmetry definitions. In both cases the distributions of asymmetry are significantly different between the BreakBRD and SFS samples. K-S tests applied to both cases estimate that the BreakBRD and control samples have only 1.7\% and 3.8\% chance of having the same distribution on the left- and right-hand sides of Figure~\ref{fig:hi_asym}, respectively. We note that the requirement for peak $S/N>5$ in this analysis was to ensure that the central \ion{H}{1} velocity is well-constrained, but the requirement is not technically necessary for the analysis using the optical velocity to define the galaxy systemic velocity. Lowering the $S/N$ requirement to $>3$ increases the sample by $\sim2$ and yields an even stronger difference in the \ion{H}{1} asymmetry distributions. We conclude that BreakBRDs are more likely to have asymmetric \ion{H}{1} profiles compared to SFS galaxies, indicating more frequent asymmetry either in their \ion{H}{1} distributions, kinematics, or both. 

\subsubsection{Kinematic Misalignment} \label{sec:manga_kin}
Kinematic quantities from integrated \ion{H}{1} spectra can be difficult to interpret given that one cannot disentangle the flux and velocity distributions. Thankfully, fourteen BreakBRD galaxies have spatially resolved optical spectroscopic data from the SDSS-IV MaNGA survey. Using this data we conduct a visual, qualitative comparison of the orientation of the gas and stellar velocity fields. 

\begin{figure*}
    \plotone{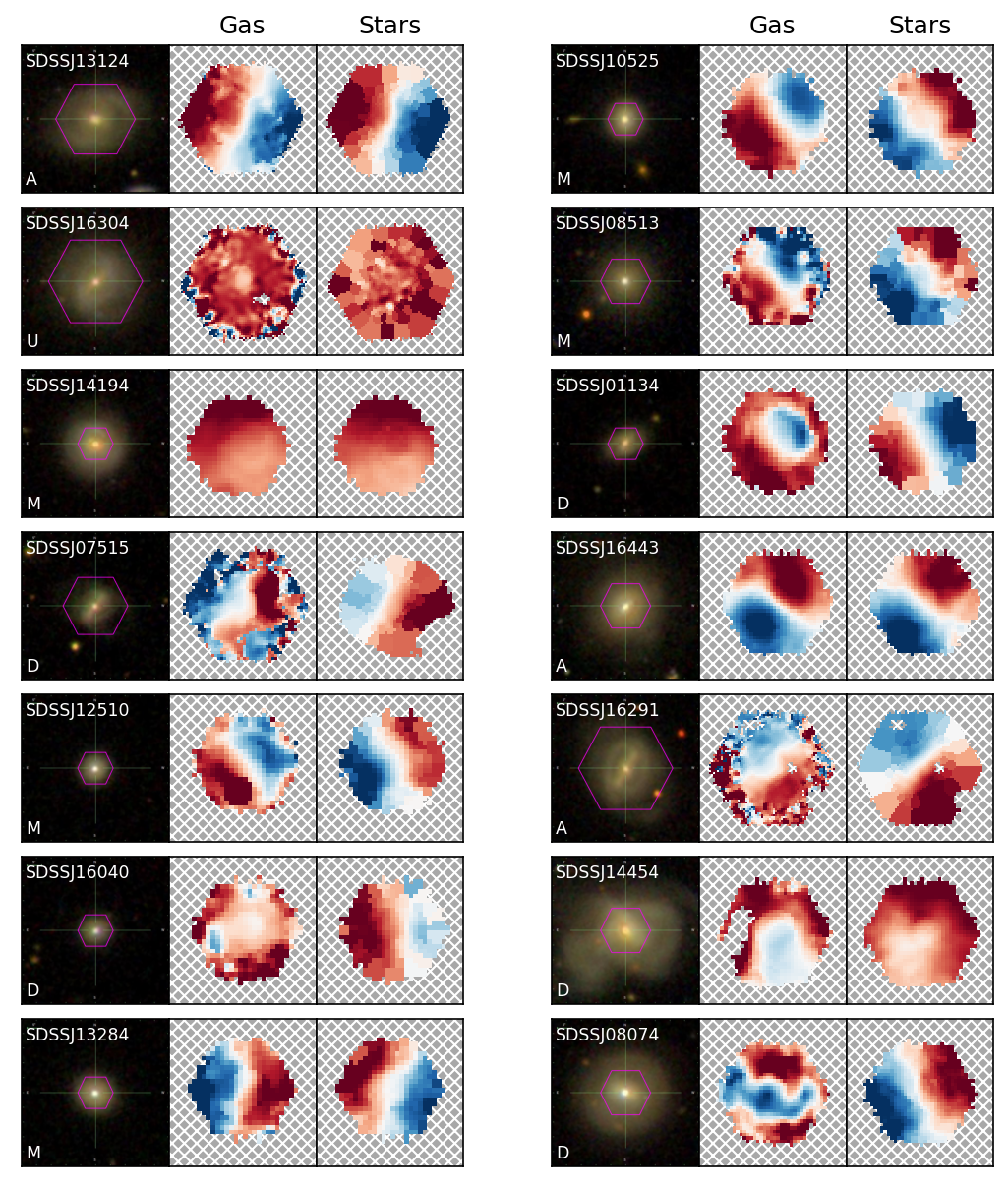}
    \caption{SDSS $gri$ images, stellar velocity fields, and gas velocity fields for the fourteen BreakBRDs with MaNGA IFU data, where blue indicates positive values and red indicates negative values. Labels in the bottom left panel of each $gri$ image indicate the classification of the gas velocity field(A: aligned, M: misaligned, D: distorted, U: uncertain). Approximately two-thirds of galaxies show either strongly misaligned or distorted gas velocity fields relative to their stellar velocity fields.}
    \label{fig:vfields}
\end{figure*}

Figure~\ref{fig:vfields} shows the $SDSS$ $gri$ image, stellar velocity field, and gas velocity field for each of the fourteen BreakBRDs with MaNGA data. Each galaxy is labeled with a visual classification for the gas/stellar velocity field: {\bf A} (aligned gas/stellar velocity fields), {\bf M} (misaligned gas/stellar velocity fields), {\bf D} (strongly distorted gas velocity field), {\bf U} (uncertain). The key difference between misaligned and distorted gas velocity fields is that misaligned cases still show clear ordered rotation, whereas distorted cases are dominated by noncircular motions whose kinematic position angle cannot easily be discerned. These classifications are all done by-eye, as strong disagreement in the gas and stellar velocity fields is often clearly visible in spatially resolved maps, but subtle differences may be missed.

In one case (8550-12703), we are unable to determine the level of agreement between the gas and stellar velocity field, possibly due to an extremely face-on configuration. Of the remaining 13, 3/13 show co-rotating velocity fields (0.13-0.36), 5/13 show clear misalignment (0.26-0.52), and 5/13 show clear distortion (0.26-0.52). All confidence intervals are 68\% and come from binomial statistics. Therefore, the majority of BreakBRDs (10/13, 68\% confidence interval (0.64-0.86) do not show simple co-rotating gas and stars. This finding, combined with the increased \ion{H}{1} asymmetry, may indicate BreakBRDs frequently host disturbed/reforming gas disks. Unfortunately, only a few galaxies contain velocity fields and $S/N > 5$ \ion{H}{1} spectra good enough to reliably measure \ion{H}{1} asymmetry, so we cannot robustly determine whether the behavior in optical velocity is correlated with the \ion{H}{1} asymmetry.

\section{Discussion} \label{sec:discussion}

In Section~\ref{sec:results}, we presented the gas properties of BreakBRD galaxies, particularly in relation to galaxies on the star-forming sequence (SFS) and Green Valley (GV), in an effort to find clues into their origin, specifically focusing on \ion{H}{1}-to-stellar mass ratio ($M_{\rm HI}/M_*$), \ion{H}{1} depletion time ($\tau_{HI}$), central gas-phase metallicity ($12+\log{O/H}$), and gas kinematics. We find that BreakBRDs have \ion{H}{1} fractions that fall between typical values for SFS and GV galaxies. In contrast, their \ion{H}{1} depletion times are on average much lower than \textit{both} SFS and GV galaxies. BreakBRDs have a broader distribution of gas-phase metallicities, one that skews towards slightly higher metallicities than the control sample. In comparison to the SFS control, they fall systematically low on the BTFR, and they have significantly higher rates of \ion{H}{1} profile asymmetry, even after accounting for different definitions of the kinematic center. The majority of galaxies with spatially resolved velocity fields show either highly disturbed internal gas motions or significant misalignment with stellar rotation. 

In the following section, we discuss how these results inform our understanding of how BreakBRDs fit into our understanding of galaxy evolution, first discussing whether BreakBRDs are truly a quenching population, and then whether their behavior is driven by merger activity.

\subsection{Are BreakBRDs on the path towards quenching?}
\label{sec:quenching}

\citet{Tuttle20} describe the BreakBRD population as not clearly fitting into a single evolutionary paradigm, e.g., they are not all satellites potentially undergoing ram-pressure stripping, or barred galaxies experiencing gas inflows. From their analysis, it was unclear whether they fell into the population of quenching galaxies, or normal galaxies undergoing a redistribution in the location of star formation (due to e.g., gas flows), or some combination of the two.

Investigations into the spatially resolved star formation properties of BreakBRDs using MaNGA data (McKay et al., in preparation) show that BreakBRDs have clearly deficient disk star formation relative to ``typical" SFS galaxies that show ``inside-out" evolution (disks bluer than bulges). The specific star formation rates (sSFRs) of the disks fall below the main-sequence values, but are not yet consistent with fully quenched galaxies. Meanwhile, BreakBRDs have bulge and global sSFRs that are generally consistent with SFS galaxies. These findings disfavor any scenario where gas has been redistributed from the disk to the bulge, which would likely drive an {\it enhancement} in central sSFR. Instead, these findings favor a scenario where galaxy-wide star formation is shutting down from the outside-in. 

Our own results show BreakBRDs have much lower gas content and \ion{H}{1} depletion times than SFS galaxies. These results imply that BreakBRDs are exhausting their gas supply. Presumably, given the outside-in cessation of star formation, the disk \ion{H}{1} has either been depleted or directly removed, with no further accretion of gas to balance the loss. While we have no spatial information on the \ion{H}{1} distribution, the simulations of \citet{Tonnesen2023} show BreakBRDs have lower than average CGM masses, consistent with them lacking gas infall to replenish their \ion{H}{1} disks. \citep{Kopenhafer2020} find that BreakBRDs in simulations have lower overall dense gas mass, but comparable dense gas within their central 2 kpc compared to typical star-forming galaxies. These studies, combined with the McKay et al. (in preparation) results, imply that the short \ion{H}{1} depletion times in BreakBRDs are the result of lower overall gas content but still nominally normal levels of SF. If instead the cold gas in BreakBRDs had been redistributed towards the center, the overall \ion{H}{1} content should not evolve significantly unless there was a major starburst, which we do not see. 

Their more rapid gas exhaustion distinguishes BreakBRDs from the general population of quenching galaxies that occupy the Green Valley. Most Green Valley galaxies show extremely long gas depletion times (see our Fig.~\ref{fig:tdep_mstars}, as well as \citealt{Janowiecki2020}), implying that although some gas may be present, it does not have the appropriate conditions to transition into stars. We see two possible reasons for why BreakBRDs have shorter depletion times. They may be in an earlier stage of quenching, where the gas supply has declined but there is still enough gas to sustain ``normal" levels of SF for now. Alternatively, BreakBRDs may have experienced events (e.g., tidal interactions) that triggered nuclear inflows enabling gas to reach densities to trigger more rapid star formation. In the first scenario, we expect BreakBRDs to have higher gas fractions than Green Valley galaxies, which they do on-average (Figure~\ref{fig:gs_mstars}). The gas fractions expected in the second scenario may depend on the mode and time elapsed since gas was perturbed and driven inwards. If BreakBRDs are the result of very recent interactions, we might expect their gas fractions to be comparable to other Green Valley galaxies. If significant time has passed since the initial inflow, BreakBRDs may have consumed enough gas in star formation to lead to {\it reduced} gas fractions compared to galaxies in the Green Valley. However, these expectations may break down if the star formation trigger was the result of a merger with another galaxy that deposited material onto the BreakBRD. It is also important to remember that the results of McKay et al. (in preparation) do not show evidence for systematically higher SFR in BreakBRDs, disfavoring a ``triggered burst" scenario. We conclude that the \ion{H}{1} fractions and SFRs more clearly support the early-stage quenching scenario.

The slight increase in nuclear metallicity observed in much of the BreakBRD sample may be another reflection of this population moving off the star forming sequence. The fundamental mass-metallicity relation states that at fixed stellar mass, galaxies with higher SFR will have lower metallicity \citep{Mannucci2010,Andrews2013}. Such behavior implies that at fixed mass, enhanced metallicity is a natural byproduct of the quenching process. While a full review of the origin of the FMR is beyond the scope of this work, an increase in the metallicity during the quenching phase may be linked to a deficit of cold gas (which we have established for the BreakBRDs) that would be sustained by fresh accretion and normally dilute metallicity. Thus, their enhanced nuclear metallicities may be a reflection of them transitioning to the quenched regime. A related result was reported by \citet{Tonnesen2023}, who found simulated BreakBRDs had elevated CGM metallicities. Based on the lower than average CGM mass and weak SFR of BreakBRDs, they argue that the most likely explanation for the enhanced metallicities is a lack of pristine gas infall.
We do note, however, that while the distribution of central BreakBRD metallicities is skewed towards higher values, it still shows a broad range.

There is an alternative explanation for at least some BreakBRDs: they are previously quenched galaxies that experienced a recent accretion event, potentially rebuilding a gas disk and reigniting star formation (see \citealt{KGB} for a discussion of such events). Such a scenario is supported by the high rate of misaligned gas/stellar disks, which imply an external gas origin. Accretion of gas may arise from mergers with a wide range of mass ratios, potentially explaining the large diversity of \ion{H}{1} fractions for BreakBRDs. In such a scenario, there may be significant debris and any newly formed gas disk may take significant time to resettle, which would in-turn explain the high \ion{H}{1} profile asymmetries. \citet{Tonnesen2023} find that the CGM around BreakBRDs is misaligned as well, but it also has low angular momentum. This combination will lead to gas to most likely enter the disk at the center, reigniting star formation there first, and consistent with our observations \citep{VanDeVoort15, Trapp2022}. It is unclear in the long run whether such BreakBRDs would move up into the star forming sequence, or down into the quenched regime. A sample of galaxies similar to BreakBRDs via their blue-centered color gradients, blue sequence E/S0s, are thought to contain a significant number of rebuilding disks, although this behavior is most common at lower stellar masses beyond those of our sample \citep{KGB}. It is also worth noting that blue-sequence E/S0s often have notably blue disks, while BreakBRDs are defined as having red disks, so it is unclear if these populations can be treated similarly. Furthermore, \citet{Kopenhafer2020} find that simulated BreakBRDs are indeed quenching, disfavoring the idea that BreakBRDs may be rejuvenated for any long period of time.

\subsection{Are BreakBRDs the result of mergers}
\label{sec:mergers}

\subsubsection{Morphological Evidence} 
\label{sec:mergers_morph}
Only a handful of BreakBRD galaxies show morphologies in their optical images clearly indicative of ongoing merger activity (e.g., SDSSJ14454 in Figure~\ref{fig:vfields}). However, blue-centered color gradients, which reflect a higher than normal ratio of central to disk SF likely due to gas inflows, are another possible indicator of past interactions \citep{Kannappan04, Gonzalez-Perez2011, Stark13}. Bars may also drive such behavior, but the bar fraction in BreakBRDs is $\sim$30-40\% \citep{Tuttle20}, typical for the general galaxy population based on SDSS imaging data \citep{Masters11}, and not high enough to explain the entire BreakBRD population. If breakBRDs are typical SFS galaxies that experienced tidal- or bar-driven gas inflows, these inflows would have to be significant enough to starve the disk of gas, yet modest enough to not drive strong central starbursts (which BreakBRDs do not have; McKay et al. in preparation). 

\begin{figure*} 
    \plottwo{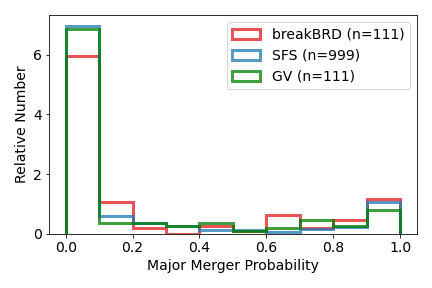}{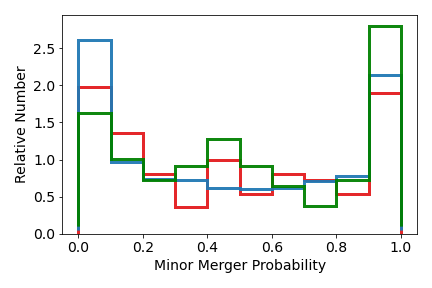}
    \caption{Probabilities of BreakBRD and control galaxies being major mergers (left) and minor mergers (right). All probabilities are based on the LDA technique of \citet{Nevin19} applied to SDSS imaging data. The minor merger probabilities are statistically the same. The major merger probability distributions are distinct, but there is no significant difference in the fraction of major mergers classified as having probability $>0.5$.}
    \label{fig:merger_prob}
\end{figure*}

To further test morphological signatures of merger activity, we employ the major and minor merger classifications from \citet{Nevin19} who combine several classical quantitative morphological parameters (e.g., image asymmetry, Gini, M20) from SDSS images into a single classifier using Linear Discriminant Analysis (LDA). The LDA technique is estimated to be sensitive to merger signatures for $>2$Gyr for both major and minor mergers. Results for the BreakBRD sample and a stellar-mass and redshift matched SFS and GV control samples  (the latter to ensure similar physical resolution) are shown in Figure~\ref{fig:merger_prob}. All samples show broad distributions in merger probability. Kolmogorov-Smirnov (K-S) and Anderson-Darling (A-D) tests are used to determine if the distributions are different.  Considering the BreakBRD and SFS control samples, we calculate K-S and A-D p-values of $2\times10^{-4}$ and $<0.001$ for the major merger probability distributions, and p-values of 0.69 and $>0.25$ for the minor merger probability distributions. Considering the BreakBRD and GV control samples, we calculate K-S and A-D p-values of 0.05 and 0.04 for the major merger probability distributions, and p-values of 0.31 and 0.20 for the minor merger probability distributions. In other words, the minor merger probabilities are effectively the same between all three samples, while the major merger probabilities are significantly different between the BreakBRD and SFS samples, and marginally different between the BreakBRD and GV samples. However, even in cases where the KS/AD tests indicate significant differences, meaningful differences in the distributions are difficult to discern by eye. If we define anything with probability $>0.5$ as a merger, then BreakBRDs have a major and minor merger rate (and 95\% confidence interval) of 0.25 (0.18-0.33) and 0.46 (0.38-0.55). Meanwhile the SFS control sample has major and minor merger rates of  0.19 (0.16-0.21) and 0.47 (0.44-0.50), and the GV control sample has major and minor merger rates of 0.22 (0.16-0.31) and 0.54 (0.45-0.62). Therefore, based on available imaging data, BreakBRDs do not appear to have a significantly higher merger fraction (at more than a $2\sigma$ level) than the general star forming galaxy population.

\subsubsection{Dynamical Evidence}
\label{sec:discussion_dynamics}

The cold gas in galaxies typically reflects the influence of mergers and interactions for $>$1Gyr after the initial encounter \citep{Holwerda11}, and therefore may be more sensitive to merger activity than analysis of optical morphology. One of our results is that BreakBRDs systematically fall below the BTFR, but the question that arises is whether this is due to them having lower than average baryonic mass or higher than average linewidth. Similar offsets below the BTFR have been seen by \citet{McGaugh2010} for observations of ultra-faint Milky Way dwarfs, and by \citet{Glowacki2020} among gas-poor galaxies in the Simba suite of simulations. Both authors cite gas-loss/quenching as a possible cause of this deviation, and \citet{McGaugh2010} correlate the offsets with tidal forces. \citet{Puech2010}, who analyzed the stellar mass TFR, also see large offsets associated with major mergers, although their scatter is more symmetric about the mean relation. If the gas disks of BreakBRDs were substantially disrupted by past or ongoing mergers, more scatter in the BTFR would be expected as the linewidth no longer traces simple circular velocity. Indeed, we find above-average \ion{H}{1} asymmetries for BreakBRDs, which could be linked to merger activity. However, interpretation of \ion{H}{1} asymmetries must be made cautiously; while there are studies that have argued that high asymmetries are associated with mergers and tidal interactions \citep{Angiras2007,vanEymeren2011,Ramirez-Moreta2018,Scott2018, Bok2019, Watts2020}, a similar breadth of literature has argued other processes, such as cold gas accretion \citep{Bournaud05, Sancisi08,Ramirez-Moreta2018} and lopsided dark matter halos \citep{Jog1997} may also play a role, especially given the existence of highly asymmetric disks in the absence of any clear companion \citep{Richter94,Haynes98,Espada11}. \citet{Watts2023} note that \ion{H}{1} asymmetries can trace disturbances that do not impact the central regions of galaxies,

It is noteworthy that the high rate of \ion{H}{1} asymmetries in BreakBRDs is coupled with a high rate (roughly two-thirds) of gas velocity fields that are distorted or completely misaligned from their stellar components. This rate is much higher than the overall galaxy population; using a broad sample of 1220 galaxies from the first year of  SDSS-IV MaNGA observations (with no additional selections made on other properties), \citet{Jin16} find that the typical rate of misalignment is a few percent on average, although the fraction depends significantly on properties like stellar mass, star formation rate, and specific star formation rate. At the typical sSFR of the BreakBRD sample ($\sim10^{-10}\,{\rm yr^{-1}}$), the rate of misaligned disks is still only a few percent. Our rate of 38\% (23\%-56\%) therefore appears significant.

Misaligned gas can \textit{only} arise if the gas arrived from outside the galaxy and has an angular momentum vector misaligned with the stars. Such gas may have arrived through cosmological accretion or the accretion of a small gas-rich companion \citep{Serra12,Lagos15}. Alternatively, a misaligned disk can result from a strong merger where the disk completely reforms on top of the stellar remnant \citep{VanDeVoort15}. Any infalling gas should not have any preferred orientation angle, which means the misaligned disks we observe do not capture all the recently formed disks because some may have ended in a co-rotating gas/star configuration, in which case our fraction of misaligned disks represent a lower limit on the total number of recently (re)formed disks. 

If the kinematic behavior of BreakBRDs is due to past merger activity, we must reconcile that with the lack of merger evidence from imaging data. \citet{Tonnesen2023} note that the combination of low CGM mass and extended misalignment in simulated BreakBRDs could lead the induced central star-formation to be very long-lived due to a slow CGM infall rate. Such behavior could explain the lack of clear optical merger signatures if they dissipate more rapidly.

We conclude that the kinematic data supports recent external gas infall in many BreakBRDs, but it is unclear whether the gas disk misalignment, asymmetries, and BTFR offsets can be directly tied to merger events.

\subsubsection{Metallicity evidence}

BreakBRDs also show a clear broadening of their metallicity distribution, one that is notably skewed towards slightly higher gas-phase metallicities compared to the control sample. If BreakBRDs are in a state of quenching (see Section~\ref{sec:quenching}), the slight increase in metallicity is expected as the population moves off the star-forming sequence (although there may be a mass dependence for this behavior; \citealt{Kumari2021}). Therefore, systematic shifts in metallicity do not necessarily point to merging activity, but given the other potential evidence (mostly dynamical), we can consider how mergers might affect and be consistent with our observed metallicity trends.

\begin{figure}
    \includegraphics[width=\columnwidth]{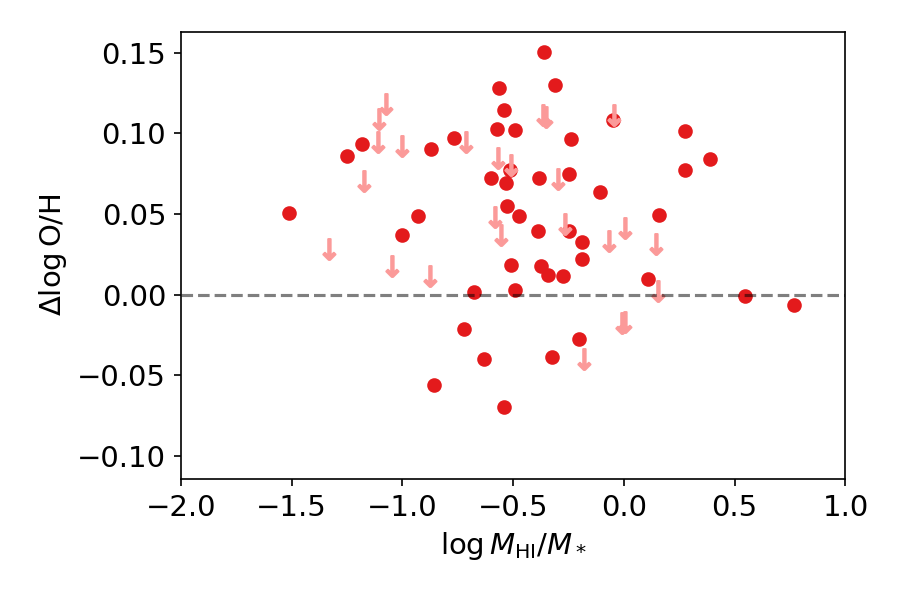}
    \caption{Residuals from the mass-metallicity relation of \citet{Marino13} versus $\log{M_{\rm HI}/M_*}$ for the BreakBRD sample. We see no positive correlation between \ion{H}{1} content and metallicity enhancement, as seen in mergers in the simulations of \citet{Torrey12}.}
     \label{fig:gs_mzr_dist}
\end{figure}

Several studies have explored the influence of merger activity on central gas-phase metallicity, finding that galaxies undergoing interactions tend to have periods with lower central metallicity caused by dilution from inflowing gas \citep{Michel-Dansac2008,Rupke10,Perez2011}. However, the timescale and stage of a merger is a major factor. For example, simulations have shown that the dilution phase from major mergers and flyby interactions is $<500\,{\rm Myr}$ in many cases, and is followed by a net enhancement in central metallicity at the moment of, or soon after, final coalescence \citep{Montuori2010,Bustamante2018}. These results suggest the metallicity enhancement we observe is still consistent with a recent merger scenario, as long as BreakBRDs are post-coalescence. \citet{Torrey12} also predict that metallicity can increase after mergers, and this metallicity offset correlates with the gas content of the merger. However, net enhancements in metallicity are mainly expected for high redshift galaxies. At $z\sim0$, nuclear metallicities are expected to generally drop after a merger, although \citet{Torrey12} predict the net dilution to be less as gas fraction increases. We have examined whether BreakBRDs show any correlation between \ion{H}{1} fraction and offset from the mass-metallicity relation of \citet{Marino13}, but find none (Figure~\ref{fig:gs_mzr_dist}). This null result may not be surprising as the range of \ion{H}{1} fractions in our data is much smaller than the simulations of \citet{Torrey12} that show a correlation. Nonetheless, this analysis does not indicate a link between higher metallicities and mergers.

\section{Conclusions} \label{sec:conclusions}

Using newly obtained deep 21cm observations from the GBT, in combination with optical spectroscopy, we have presented an analysis of the gas properties of BreakBRD galaxies, a population characterized by their star forming bulges and weakly or non-star-forming disks. We compare the properties of BreakBRDs with control samples of galaxies on the Star Forming Sequence (SFS) and Green Valley (GV). Our key results are:
\begin{itemize}
    \item BreakBRDs have a wide range of gas fractions at fixed stellar mass, but their distribution has a lower mean value than typical SFS galaxies, and a higher mean value than GV galaxies.
    \item BreakBRDs have significantly lower average \ion{H}{1} depletion times compared to SFS galaxies, in contrast with GV galaxies that tend to have much {\it longer} depletion times than SFS galaxies.
    \item The distribution of nuclear gas-phase metallicity is notably broader than that for the main star-forming population, and is skewed towards higher values. It is similar to GV galaxies, among the subset where metallicity can be estimated.
    \item BreakBRDs falls systematically low on the Baryonic Tuller-Fisher Relation (BTFR) relative to SFS galaxies.
    \item BreakBRDs have high \ion{H}{1} asymmetries and 2/3 of those with MaNGA IFU spectroscopy show misaligned or disturbed gas velocity fields. However, it is unclear from our data set whether the optical and \ion{H}{1} kinematic properties are directly correlated.
\end{itemize}

A unified picture of BreakBRD evolution from these data remains difficult to determine, and it is still possible that BreakBRDs represent a variety of ongoing evolutionary processes. However, their gas deficiency coupled with resolved star formation properties is generally consistent with a population transitioning into the quenched regime, but perhaps at an early stage. Metallicities skewed towards higher values at fixed mass compared to galaxies on the star-forming sequence may also support this scenario, with the caveat that the metallicity distribution of BreakBRDs is quite broad, with some being metal {\it deficient}. Frequent high \ion{H}{1} asymmetry and star/gas velocity misalignment could indicate that BreakBRDs have been subject of gas infall and/or tidal interactions, although imaging data generally do not clearly indicate recent merger activity. It is also possible that a subset of BreakBRDs represent previously quenched galaxies with recent gas infall that rejuvenated their central star formation. 

Resolving the large-scale gas distribution may provide a clearer understanding of BreakBRD evolution. Spatially resolved \ion{H}{1} is an ideal tracer of recent interactions, and will more clearly indicate whether there are inflows or resettling disks, and they will enable us to assess whether \ion{H}{1} asymmetries and disturbed/counter-rotating gas are related to each other. Such observations will also allow us to determine whether the \ion{H}{1} is significantly truncated beyond the bulges of BreakBRDs, or present but with densities or dynamics that prevent it from contributing to star formation. Furthermore, it will enable us to measure disk angular momentum more accurately and determine whether metallicity enhancements and gas-star misalignment are associated with high and low angular momentum, respectively, as has been reported by \citet{Tonnesen2023}. New Karl G. Jansky Very Large Array observations are underway for a pilot sample of BreakBRDs and will allow us to address these questions.  Additionally, while an initial assessment by \citet{Tuttle20} demonstrated that BreakBRDs are found in a wide range of environments, a more detailed analysis of environment (including multiple metrics) and its relationship to BreakBRD properties would be beneficial. Such a study is planned as future work.

\begin{acknowledgments}

We thank the anonymous referee for their careful review which helped improve this paper. This work was made possible through NSF grant 1813462 and STScI DDRF Grant D0001.82511. The authors thank Rebecca Nevin for sharing her imaging-based merger probabilities for our sample.

The authors acknowledge the work of the entire ALFALFA collaboration who have contributed to the many aspects of the survey over the years.

The Green Bank Observatory is a facility of the National Science Foundation operated under cooperative agreement by Associated Universities, Inc.

Funding for the Sloan Digital Sky 
Survey IV has been provided by the 
Alfred P. Sloan Foundation, the U.S. 
Department of Energy Office of 
Science, and the Participating 
Institutions. 

SDSS-IV acknowledges support and 
resources from the Center for High 
Performance Computing  at the 
University of Utah. The SDSS 
website is www.sdss4.org.

SDSS-IV is managed by the 
Astrophysical Research Consortium 
for the Participating Institutions 
of the SDSS Collaboration including 
the Brazilian Participation Group, 
the Carnegie Institution for Science, 
Carnegie Mellon University, Center for 
Astrophysics | Harvard \& 
Smithsonian, the Chilean Participation 
Group, the French Participation Group, 
Instituto de Astrof\'isica de 
Canarias, The Johns Hopkins 
University, Kavli Institute for the 
Physics and Mathematics of the 
Universe (IPMU) / University of 
Tokyo, the Korean Participation Group, 
Lawrence Berkeley National Laboratory, 
Leibniz Institut f\"ur Astrophysik 
Potsdam (AIP),  Max-Planck-Institut 
f\"ur Astronomie (MPIA Heidelberg), 
Max-Planck-Institut f\"ur 
Astrophysik (MPA Garching), 
Max-Planck-Institut f\"ur 
Extraterrestrische Physik (MPE), 
National Astronomical Observatories of 
China, New Mexico State University, 
New York University, University of 
Notre Dame, Observat\'ario 
Nacional / MCTI, The Ohio State 
University, Pennsylvania State 
University, Shanghai 
Astronomical Observatory, United 
Kingdom Participation Group, 
Universidad Nacional Aut\'onoma 
de M\'exico, University of Arizona, 
University of Colorado Boulder, 
University of Oxford, University of 
Portsmouth, University of Utah, 
University of Virginia, University 
of Washington, University of 
Wisconsin, Vanderbilt University, 
and Yale University.

\end{acknowledgments}

%

\vspace{5mm}
\facilities{SDSS, GBT}


\software{GBTIDL, lifelines}
\citep{Davidson-Pilon2019}

\bibliography{mybib}{}

\begin{thebibliography}{}
\expandafter\ifx\csname natexlab\endcsname\relax\def\natexlab#1{#1}\fi
\providecommand{\url}[1]{\href{#1}{#1}}
\providecommand{\dodoi}[1]{doi:~\href{http://doi.org/#1}{\nolinkurl{#1}}}
\providecommand{\doeprint}[1]{\href{http://ascl.net/#1}{\nolinkurl{http://ascl.net/#1}}}
\providecommand{\doarXiv}[1]{\href{https://arxiv.org/abs/#1}{\nolinkurl{https://arxiv.org/abs/#1}}}

\bibitem[{{Andrews} \& {Martini}(2013)}]{Andrews2013}
{Andrews}, B.~H., \& {Martini}, P. 2013, \apj, 765, 140,
  \dodoi{10.1088/0004-637X/765/2/140}

\bibitem[{{Angiras} {et~al.}(2007){Angiras}, {Jog}, {Dwarakanath}, \&
  {Verheijen}}]{Angiras2007}
{Angiras}, R.~A., {Jog}, C.~J., {Dwarakanath}, K.~S., \& {Verheijen}, M.~A.~W.
  2007, \mnras, 378, 276, \dodoi{10.1111/j.1365-2966.2007.11779.x}

\bibitem[{{Baldwin} {et~al.}(1981){Baldwin}, {Phillips}, \&
  {Terlevich}}]{Baldwin81}
{Baldwin}, J.~A., {Phillips}, M.~M., \& {Terlevich}, R. 1981, \pasp, 93, 5,
  \dodoi{10.1086/130766}

\bibitem[{{Bell} {et~al.}(2004){Bell}, {Wolf}, {Meisenheimer}, {Rix}, {Borch},
  {Dye}, {Kleinheinrich}, {Wisotzki}, \& {McIntosh}}]{Bell2004}
{Bell}, E.~F., {Wolf}, C., {Meisenheimer}, K., {et~al.} 2004, \apj, 608, 752,
  \dodoi{10.1086/420778}

\bibitem[{{Boissier} \& {Prantzos}(1999)}]{Boissier1999}
{Boissier}, S., \& {Prantzos}, N. 1999, \mnras, 307, 857,
  \dodoi{10.1046/j.1365-8711.1999.02699.x}

\bibitem[{{Bok} {et~al.}(2019){Bok}, {Blyth}, {Gilbank}, \& {Elson}}]{Bok2019}
{Bok}, J., {Blyth}, S.~L., {Gilbank}, D.~G., \& {Elson}, E.~C. 2019, \mnras,
  484, 582, \dodoi{10.1093/mnras/sty3448}

\bibitem[{{Bournaud} {et~al.}(2005){Bournaud}, {Combes}, \&
  {Semelin}}]{Bournaud05}
{Bournaud}, F., {Combes}, F., \& {Semelin}, B. 2005, \mnras, 364, L18,
  \dodoi{10.1111/j.1745-3933.2005.00096.x}

\bibitem[{{Bundy} {et~al.}(2015){Bundy}, {Bershady}, {Law}, {Yan}, {Drory},
  {MacDonald}, {Wake}, {Cherinka}, {S{\'a}nchez-Gallego}, {Weijmans}, {Thomas},
  {Tremonti}, {Masters}, {Coccato}, {Diamond-Stanic}, {Arag{\'o}n-Salamanca},
  {Avila-Reese}, {Badenes}, {Falc{\'o}n-Barroso}, {Belfiore}, {Bizyaev},
  {Blanc}, {Bland-Hawthorn}, {Blanton}, {Brownstein}, {Byler}, {Cappellari},
  {Conroy}, {Dutton}, {Emsellem}, {Etherington}, {Frinchaboy}, {Fu}, {Gunn},
  {Harding}, {Johnston}, {Kauffmann}, {Kinemuchi}, {Klaene}, {Knapen},
  {Leauthaud}, {Li}, {Lin}, {Maiolino}, {Malanushenko}, {Malanushenko}, {Mao},
  {Maraston}, {McDermid}, {Merrifield}, {Nichol}, {Oravetz}, {Pan}, {Parejko},
  {Sanchez}, {Schlegel}, {Simmons}, {Steele}, {Steinmetz}, {Thanjavur},
  {Thompson}, {Tinker}, {van den Bosch}, {Westfall}, {Wilkinson}, {Wright},
  {Xiao}, \& {Zhang}}]{Bundy15}
{Bundy}, K., {Bershady}, M.~A., {Law}, D.~R., {et~al.} 2015, \apj, 798, 7,
  \dodoi{10.1088/0004-637X/798/1/7}

\bibitem[{{Bustamante} {et~al.}(2018){Bustamante}, {Sparre}, {Springel}, \&
  {Grand}}]{Bustamante2018}
{Bustamante}, S., {Sparre}, M., {Springel}, V., \& {Grand}, R. J.~J. 2018,
  \mnras, 479, 3381, \dodoi{10.1093/mnras/sty1692}

\bibitem[{{Cano-D{\'\i}az} {et~al.}(2019){Cano-D{\'\i}az}, {{\'A}vila-Reese},
  {S{\'a}nchez}, {Hern{\'a}ndez-Toledo}, {Rodr{\'\i}guez-Puebla}, {Boquien}, \&
  {Ibarra-Medel}}]{Cano-Diaz19}
{Cano-D{\'\i}az}, M., {{\'A}vila-Reese}, V., {S{\'a}nchez}, S.~F., {et~al.}
  2019, \mnras, 488, 3929, \dodoi{10.1093/mnras/stz1894}

\bibitem[{{Carles} {et~al.}(2016){Carles}, {Martel}, {Ellison}, \&
  {Kawata}}]{Carles2016}
{Carles}, C., {Martel}, H., {Ellison}, S.~L., \& {Kawata}, D. 2016, \mnras,
  463, 1074, \dodoi{10.1093/mnras/stw2056}

\bibitem[{{Chang} {et~al.}(2015){Chang}, {van der Wel}, {da Cunha}, \&
  {Rix}}]{Chang15}
{Chang}, Y.-Y., {van der Wel}, A., {da Cunha}, E., \& {Rix}, H.-W. 2015, \apjs,
  219, 8, \dodoi{10.1088/0067-0049/219/1/8}

\bibitem[{{Chiappini} {et~al.}(1997){Chiappini}, {Matteucci}, \&
  {Gratton}}]{Chiappini1997}
{Chiappini}, C., {Matteucci}, F., \& {Gratton}, R. 1997, \apj, 477, 765,
  \dodoi{10.1086/303726}

\bibitem[{Davidson-Pilon(2019)}]{Davidson-Pilon2019}
Davidson-Pilon, C. 2019, Journal of Open Source Software, 4, 1317,
  \dodoi{10.21105/joss.01317}

\bibitem[{{Ellison} {et~al.}(2011){Ellison}, {Nair}, {Patton}, {Scudder},
  {Mendel}, \& {Simard}}]{Ellison11}
{Ellison}, S.~L., {Nair}, P., {Patton}, D.~R., {et~al.} 2011, \mnras, 416,
  2182, \dodoi{10.1111/j.1365-2966.2011.19195.x}

\bibitem[{{Erroz-Ferrer} {et~al.}(2019){Erroz-Ferrer}, {Carollo}, {den Brok},
  {Onodera}, {Brinchmann}, {Marino}, {Monreal-Ibero}, {Schaye}, {Woo},
  {Cibinel}, {Debattista}, {Inami}, {Maseda}, {Richard}, {Tacchella}, \&
  {Wisotzki}}]{ErrozFerrer2019}
{Erroz-Ferrer}, S., {Carollo}, C.~M., {den Brok}, M., {et~al.} 2019, \mnras,
  484, 5009, \dodoi{10.1093/mnras/stz194}

\bibitem[{{Espada}(2011)}]{Espada11}
{Espada}, D. 2011, \aap, 532, A117, \dodoi{10.1051/0004-6361/201016117}

\bibitem[{{Faber} {et~al.}(2007){Faber}, {Willmer}, {Wolf}, {Koo}, {Weiner},
  {Newman}, {Im}, {Coil}, {Conroy}, {Cooper}, {Davis}, {Finkbeiner}, {Gerke},
  {Gebhardt}, {Groth}, {Guhathakurta}, {Harker}, {Kaiser}, {Kassin},
  {Kleinheinrich}, {Konidaris}, {Kron}, {Lin}, {Luppino}, {Madgwick},
  {Meisenheimer}, {Noeske}, {Phillips}, {Sarajedini}, {Schiavon}, {Simard},
  {Szalay}, {Vogt}, \& {Yan}}]{Faber2007}
{Faber}, S.~M., {Willmer}, C.~N.~A., {Wolf}, C., {et~al.} 2007, \apj, 665, 265,
  \dodoi{10.1086/519294}

\bibitem[{{Glowacki} {et~al.}(2020){Glowacki}, {Elson}, \&
  {Dav{\'e}}}]{Glowacki2020}
{Glowacki}, M., {Elson}, E., \& {Dav{\'e}}, R. 2020, \mnras, 498, 3687,
  \dodoi{10.1093/mnras/staa2616}

\bibitem[{{Goddy} {et~al.}(2020){Goddy}, {Stark}, \& {Masters}}]{Goddy20}
{Goddy}, J., {Stark}, D.~V., \& {Masters}, K.~L. 2020, Research Notes of the
  American Astronomical Society, 4, 3, \dodoi{10.3847/2515-5172/ab66bd}

\bibitem[{{Goddy} {et~al.}(2023){Goddy}, {Stark}, {Masters}, {Bundy}, {Drory},
  \& {Law}}]{Goddy2023}
{Goddy}, J.~S., {Stark}, D.~V., {Masters}, K.~L., {et~al.} 2023, \mnras, 520,
  3895, \dodoi{10.1093/mnras/stad298}

\bibitem[{{Gonzalez-Perez} {et~al.}(2011){Gonzalez-Perez}, {Castander}, \&
  {Kauffmann}}]{Gonzalez-Perez2011}
{Gonzalez-Perez}, V., {Castander}, F.~J., \& {Kauffmann}, G. 2011, \mnras, 411,
  1151, \dodoi{10.1111/j.1365-2966.2010.17744.x}

\bibitem[{{Haynes} {et~al.}(1998){Haynes}, {van Zee}, {Hogg}, {Roberts}, \&
  {Maddalena}}]{Haynes98}
{Haynes}, M.~P., {van Zee}, L., {Hogg}, D.~E., {Roberts}, M.~S., \&
  {Maddalena}, R.~J. 1998, \aj, 115, 62, \dodoi{10.1086/300166}

\bibitem[{{Haynes} {et~al.}(2018){Haynes}, {Giovanelli}, {Kent}, {Adams},
  {Balonek}, {Craig}, {Fertig}, {Finn}, {Giovanardi}, {Hallenbeck}, {Hess},
  {Hoffman}, {Huang}, {Jones}, {Koopmann}, {Kornreich}, {Leisman}, {Miller},
  {Moorman}, {O'Connor}, {O'Donoghue}, {Papastergis}, {Troischt}, {Stark}, \&
  {Xiao}}]{Haynes18}
{Haynes}, M.~P., {Giovanelli}, R., {Kent}, B.~R., {et~al.} 2018, \apj, 861, 49,
  \dodoi{10.3847/1538-4357/aac956}

\bibitem[{{Holwerda} {et~al.}(2011){Holwerda}, {Pirzkal}, {Cox}, {de Blok},
  {Weniger}, {Bouchard}, {Blyth}, \& {van der Heyden}}]{Holwerda11}
{Holwerda}, B.~W., {Pirzkal}, N., {Cox}, T.~J., {et~al.} 2011, \mnras, 416,
  2426, \dodoi{10.1111/j.1365-2966.2011.18940.x}

\bibitem[{{Janowiecki} {et~al.}(2020){Janowiecki}, {Catinella}, {Cortese},
  {Saintonge}, \& {Wang}}]{Janowiecki2020}
{Janowiecki}, S., {Catinella}, B., {Cortese}, L., {Saintonge}, A., \& {Wang},
  J. 2020, \mnras, 493, 1982, \dodoi{10.1093/mnras/staa178}

\bibitem[{{Jin} {et~al.}(2016){Jin}, {Chen}, {Shi}, {Tremonti}, {Bershady},
  {Merrifield}, {Emsellem}, {Fu}, {Wake}, {Bundy}, {Lin}, {Argudo-Fernandez},
  {Huang}, {Stark}, {Storchi-Bergmann}, {Bizyaev}, {Brownstein}, {Chisholm},
  {Guo}, {Hao}, {Hu}, {Li}, {Li}, {Masters}, {Malanushenko}, {Pan}, {Riffel},
  {Roman-Lopes}, {Simmons}, {Thomas}, {Wang}, {Westfall}, \& {Yan}}]{Jin16}
{Jin}, Y., {Chen}, Y., {Shi}, Y., {et~al.} 2016, \mnras, 463, 913,
  \dodoi{10.1093/mnras/stw2055}

\bibitem[{{Jog}(1997)}]{Jog1997}
{Jog}, C.~J. 1997, \apj, 488, 642, \dodoi{10.1086/304721}

\bibitem[{{Kannappan}(2004)}]{Kannappan04}
{Kannappan}, S.~J. 2004, \apjl, 611, L89, \dodoi{10.1086/423785}

\bibitem[{{Kannappan} \& {Gawiser}(2007)}]{Kannappan07}
{Kannappan}, S.~J., \& {Gawiser}, E. 2007, \apjl, 657, L5,
  \dodoi{10.1086/512974}

\bibitem[{{Kannappan} {et~al.}(2009){Kannappan}, {Guie}, \& {Baker}}]{KGB}
{Kannappan}, S.~J., {Guie}, J.~M., \& {Baker}, A.~J. 2009, \aj, 138, 579,
  \dodoi{10.1088/0004-6256/138/2/579}

\bibitem[{{Kewley} {et~al.}(2006){Kewley}, {Groves}, {Kauffmann}, \&
  {Heckman}}]{Kewley06b}
{Kewley}, L.~J., {Groves}, B., {Kauffmann}, G., \& {Heckman}, T. 2006, \mnras,
  372, 961, \dodoi{10.1111/j.1365-2966.2006.10859.x}

\bibitem[{{Kolcu} {et~al.}(2022){Kolcu}, {Crossett}, {Bellhouse}, \&
  {McGee}}]{Kolcu2022}
{Kolcu}, T., {Crossett}, J.~P., {Bellhouse}, C., \& {McGee}, S. 2022, \mnras,
  515, 5877, \dodoi{10.1093/mnras/stac2177}

\bibitem[{{Koopmann} \& {Kenney}(2004{\natexlab{a}})}]{Koopmann2004a}
{Koopmann}, R.~A., \& {Kenney}, J. D.~P. 2004{\natexlab{a}}, \apj, 613, 866,
  \dodoi{10.1086/423191}

\bibitem[{{Koopmann} \& {Kenney}(2004{\natexlab{b}})}]{Koopmann2004b}
---. 2004{\natexlab{b}}, \apj, 613, 851, \dodoi{10.1086/423190}

\bibitem[{{Kopenhafer} {et~al.}(2020){Kopenhafer}, {Starkenburg}, {Tonnesen},
  \& {Tuttle}}]{Kopenhafer2020}
{Kopenhafer}, C., {Starkenburg}, T.~K., {Tonnesen}, S., \& {Tuttle}, S. 2020,
  \apj, 903, 143, \dodoi{10.3847/1538-4357/abbd9c}

\bibitem[{{Kumari} {et~al.}(2021){Kumari}, {Maiolino}, {Trussler}, {Mannucci},
  {Cresci}, {Curti}, {Marconi}, \& {Belfiore}}]{Kumari2021}
{Kumari}, N., {Maiolino}, R., {Trussler}, J., {et~al.} 2021, \aap, 656, A140,
  \dodoi{10.1051/0004-6361/202140757}

\bibitem[{{Lackner} \& {Gunn}(2012)}]{Lackner2012}
{Lackner}, C.~N., \& {Gunn}, J.~E. 2012, \mnras, 421, 2277,
  \dodoi{10.1111/j.1365-2966.2012.20450.x}

\bibitem[{{Lagos} {et~al.}(2015){Lagos}, {Padilla}, {Davis}, {Lacey}, {Baugh},
  {Gonzalez-Perez}, {Zwaan}, \& {Contreras}}]{Lagos15}
{Lagos}, C.~d.~P., {Padilla}, N.~D., {Davis}, T.~A., {et~al.} 2015, \mnras,
  448, 1271, \dodoi{10.1093/mnras/stu2763}

\bibitem[{{Li} {et~al.}(2015){Li}, {Wang}, {Lin}, {Bershady}, {Bundy},
  {Tremonti}, {Xiao}, {Yan}, {Bizyaev}, {Blanton}, {Cales}, {Cherinka},
  {Cheung}, {Drory}, {Emsellem}, {Fu}, {Gelfand}, {Law}, {Lin}, {MacDonald},
  {Maraston}, {Masters}, {Merrifield}, {Pan}, {S{\'a}nchez}, {Schneider},
  {Thomas}, {Wake}, {Wang}, {Weijmans}, {Wilkinson}, {Yoachim}, {Zhang}, \&
  {Zheng}}]{Li2015}
{Li}, C., {Wang}, E., {Lin}, L., {et~al.} 2015, \apj, 804, 125,
  \dodoi{10.1088/0004-637X/804/2/125}

\bibitem[{{Lin} {et~al.}(2019){Lin}, {Hsieh}, {Pan}, {Rembold}, {S{\'a}nchez},
  {Argudo-Fern{\'a}ndez}, {Rowlands}, {Belfiore}, {Bizyaev}, {Lacerna},
  {Riffel}, {Rong}, {Yuan}, {Drory}, {Maiolino}, \& {Wilcots}}]{Lin2019}
{Lin}, L., {Hsieh}, B.-C., {Pan}, H.-A., {et~al.} 2019, \apj, 872, 50,
  \dodoi{10.3847/1538-4357/aafa84}

\bibitem[{{L{\'o}pez Fern{\'a}ndez} {et~al.}(2018){L{\'o}pez Fern{\'a}ndez},
  {Gonz{\'a}lez Delgado}, {P{\'e}rez}, {Garc{\'\i}a-Benito}, {Cid Fernandes},
  {Schoenell}, {S{\'a}nchez}, {Gallazzi}, {S{\'a}nchez-Bl{\'a}zquez}, {Vale
  Asari}, \& {Walcher}}]{LopezFernandez2018}
{L{\'o}pez Fern{\'a}ndez}, R., {Gonz{\'a}lez Delgado}, R.~M., {P{\'e}rez}, E.,
  {et~al.} 2018, \aap, 615, A27, \dodoi{10.1051/0004-6361/201732358}

\bibitem[{{Mannucci} {et~al.}(2010){Mannucci}, {Cresci}, {Maiolino}, {Marconi},
  \& {Gnerucci}}]{Mannucci2010}
{Mannucci}, F., {Cresci}, G., {Maiolino}, R., {Marconi}, A., \& {Gnerucci}, A.
  2010, \mnras, 408, 2115, \dodoi{10.1111/j.1365-2966.2010.17291.x}

\bibitem[{{Marino} {et~al.}(2013){Marino}, {Rosales-Ortega}, {S{\'a}nchez},
  {Gil de Paz}, {V{\'{\i}}lchez}, {Miralles-Caballero}, {Kehrig},
  {P{\'e}rez-Montero}, {Stanishev}, {Iglesias-P{\'a}ramo}, {D{\'{\i}}az},
  {Castillo-Morales}, {Kennicutt}, {L{\'o}pez-S{\'a}nchez}, {Galbany},
  {Garc{\'{\i}}a-Benito}, {Mast}, {Mendez-Abreu}, {Monreal-Ibero}, {Husemann},
  {Walcher}, {Garc{\'{\i}}a-Lorenzo}, {Masegosa}, {Del Olmo Orozco},
  {Mour{\~a}o}, {Ziegler}, {Moll{\'a}}, {Papaderos},
  {S{\'a}nchez-Bl{\'a}zquez}, {Gonz{\'a}lez Delgado}, {Falc{\'o}n-Barroso},
  {Roth}, {van de Ven}, \& {Califa Team}}]{Marino13}
{Marino}, R.~A., {Rosales-Ortega}, F.~F., {S{\'a}nchez}, S.~F., {et~al.} 2013,
  \aap, 559, A114, \dodoi{10.1051/0004-6361/201321956}

\bibitem[{{Martig} {et~al.}(2009){Martig}, {Bournaud}, {Teyssier}, \&
  {Dekel}}]{Martig2009}
{Martig}, M., {Bournaud}, F., {Teyssier}, R., \& {Dekel}, A. 2009, \apj, 707,
  250, \dodoi{10.1088/0004-637X/707/1/250}

\bibitem[{{Masters} {et~al.}(2010){Masters}, {Mosleh}, {Romer}, {Nichol},
  {Bamford}, {Schawinski}, {Lintott}, {Andreescu}, {Campbell}, {Crowcroft},
  {Doyle}, {Edmondson}, {Murray}, {Raddick}, {Slosar}, {Szalay}, \&
  {Vandenberg}}]{Masters10}
{Masters}, K.~L., {Mosleh}, M., {Romer}, A.~K., {et~al.} 2010, \mnras, 405,
  783, \dodoi{10.1111/j.1365-2966.2010.16503.x}

\bibitem[{{Masters} {et~al.}(2011){Masters}, {Nichol}, {Hoyle}, {Lintott},
  {Bamford}, {Edmondson}, {Fortson}, {Keel}, {Schawinski}, {Smith}, \&
  {Thomas}}]{Masters11}
{Masters}, K.~L., {Nichol}, R.~C., {Hoyle}, B., {et~al.} 2011, \mnras, 411,
  2026, \dodoi{10.1111/j.1365-2966.2010.17834.x}

\bibitem[{{Masters} {et~al.}(2019){Masters}, {Stark}, {Pace}, {Phipps},
  {Rujopakarn}, {Samanso}, {Harrington}, {S{\'a}nchez-Gallego}, {Avila-Reese},
  {Bershady}, {Cherinka}, {Fielder}, {Finnegan}, {Riffel}, {Rowlands},
  {Shamsi}, {Newnham}, {Weijmans}, \& {Witherspoon}}]{Masters19}
{Masters}, K.~L., {Stark}, D.~V., {Pace}, Z.~J., {et~al.} 2019, \mnras, 488,
  3396, \dodoi{10.1093/mnras/stz1889}

\bibitem[{{McGaugh} {et~al.}(2000){McGaugh}, {Schombert}, {Bothun}, \& {de
  Blok}}]{McGaugh00}
{McGaugh}, S.~S., {Schombert}, J.~M., {Bothun}, G.~D., \& {de Blok}, W.~J.~G.
  2000, \apjl, 533, L99, \dodoi{10.1086/312628}

\bibitem[{{McGaugh} \& {Wolf}(2010)}]{McGaugh2010}
{McGaugh}, S.~S., \& {Wolf}, J. 2010, \apj, 722, 248,
  \dodoi{10.1088/0004-637X/722/1/248}

\bibitem[{{Michel-Dansac} {et~al.}(2008){Michel-Dansac}, {Lambas}, {Alonso}, \&
  {Tissera}}]{Michel-Dansac2008}
{Michel-Dansac}, L., {Lambas}, D.~G., {Alonso}, M.~S., \& {Tissera}, P. 2008,
  \mnras, 386, L82, \dodoi{10.1111/j.1745-3933.2008.00466.x}

\bibitem[{{Montuori} {et~al.}(2010){Montuori}, {Di Matteo}, {Lehnert},
  {Combes}, \& {Semelin}}]{Montuori2010}
{Montuori}, M., {Di Matteo}, P., {Lehnert}, M.~D., {Combes}, F., \& {Semelin},
  B. 2010, \aap, 518, A56, \dodoi{10.1051/0004-6361/201014304}

\bibitem[{{Moreno} {et~al.}(2015){Moreno}, {Torrey}, {Ellison}, {Patton},
  {Bluck}, {Bansal}, \& {Hernquist}}]{Moreno2015}
{Moreno}, J., {Torrey}, P., {Ellison}, S.~L., {et~al.} 2015, \mnras, 448, 1107,
  \dodoi{10.1093/mnras/stv094}

\bibitem[{{Nevin} {et~al.}(2019){Nevin}, {Blecha}, {Comerford}, \&
  {Greene}}]{Nevin19}
{Nevin}, R., {Blecha}, L., {Comerford}, J., \& {Greene}, J. 2019, \apj, 872,
  76, \dodoi{10.3847/1538-4357/aafd34}

\bibitem[{{Ott} {et~al.}(1994){Ott}, {Witzel}, {Quirrenbach}, {Krichbaum},
  {Standke}, {Schalinski}, \& {Hummel}}]{Ott1994}
{Ott}, M., {Witzel}, A., {Quirrenbach}, A., {et~al.} 1994, \aap, 284, 331

\bibitem[{{Peng} {et~al.}(2010){Peng}, {Lilly}, {Kova{\v{c}}}, {Bolzonella},
  {Pozzetti}, {Renzini}, {Zamorani}, {Ilbert}, {Knobel}, {Iovino}, {Maier},
  {Cucciati}, {Tasca}, {Carollo}, {Silverman}, {Kampczyk}, {de Ravel},
  {Sanders}, {Scoville}, {Contini}, {Mainieri}, {Scodeggio}, {Kneib}, {Le
  F{\`e}vre}, {Bardelli}, {Bongiorno}, {Caputi}, {Coppa}, {de la Torre},
  {Franzetti}, {Garilli}, {Lamareille}, {Le Borgne}, {Le Brun}, {Mignoli},
  {Perez Montero}, {Pello}, {Ricciardelli}, {Tanaka}, {Tresse}, {Vergani},
  {Welikala}, {Zucca}, {Oesch}, {Abbas}, {Barnes}, {Bordoloi}, {Bottini},
  {Cappi}, {Cassata}, {Cimatti}, {Fumana}, {Hasinger}, {Koekemoer},
  {Leauthaud}, {Maccagni}, {Marinoni}, {McCracken}, {Memeo}, {Meneux}, {Nair},
  {Porciani}, {Presotto}, \& {Scaramella}}]{Peng2010}
{Peng}, Y.-j., {Lilly}, S.~J., {Kova{\v{c}}}, K., {et~al.} 2010, \apj, 721,
  193, \dodoi{10.1088/0004-637X/721/1/193}

\bibitem[{{P{\'e}rez} {et~al.}(2013){P{\'e}rez}, {Cid Fernandes}, {Gonz{\'a}lez
  Delgado}, {Garc{\'\i}a-Benito}, {S{\'a}nchez}, {Husemann}, {Mast},
  {Rod{\'o}n}, {Kupko}, {Backsmann}, {de Amorim}, {van de Ven}, {Walcher},
  {Wisotzki}, {Cortijo-Ferrero}, \& {CALIFA Collaboration}}]{Perez2013}
{P{\'e}rez}, E., {Cid Fernandes}, R., {Gonz{\'a}lez Delgado}, R.~M., {et~al.}
  2013, \apjl, 764, L1, \dodoi{10.1088/2041-8205/764/1/L1}

\bibitem[{{Perez} {et~al.}(2011){Perez}, {Michel-Dansac}, \&
  {Tissera}}]{Perez2011}
{Perez}, J., {Michel-Dansac}, L., \& {Tissera}, P.~B. 2011, \mnras, 417, 580,
  \dodoi{10.1111/j.1365-2966.2011.19300.x}

\bibitem[{{Perley} \& {Butler}(2017)}]{Perley2017}
{Perley}, R.~A., \& {Butler}, B.~J. 2017, \apjs, 230, 7,
  \dodoi{10.3847/1538-4365/aa6df9}

\bibitem[{{Puech} {et~al.}(2010){Puech}, {Hammer}, {Flores}, {Delgado-Serrano},
  {Rodrigues}, \& {Yang}}]{Puech2010}
{Puech}, M., {Hammer}, F., {Flores}, H., {et~al.} 2010, \aap, 510, A68,
  \dodoi{10.1051/0004-6361/200912081}

\bibitem[{{Ram{\'\i}rez-Moreta} {et~al.}(2018){Ram{\'\i}rez-Moreta},
  {Verdes-Montenegro}, {Blasco-Herrera}, {Leon}, {Venhola}, {Yun}, {Peris},
  {Peletier}, {Verdoes Kleijn}, {Unda-Sanzana}, {Espada}, {Bosma},
  {Athanassoula}, {Argudo-Fern{\'a}ndez}, {Sabater}, {Mu{\~n}oz-Mateos},
  {Jones}, {Huchtmeier}, {Ruiz}, {Iglesias-P{\'a}ramo},
  {Fern{\'a}ndez-Lorenzo}, {Beckman}, {S{\'a}nchez-Exp{\'o}sito}, \&
  {Garrido}}]{Ramirez-Moreta2018}
{Ram{\'\i}rez-Moreta}, P., {Verdes-Montenegro}, L., {Blasco-Herrera}, J.,
  {et~al.} 2018, \aap, 619, A163, \dodoi{10.1051/0004-6361/201833333}

\bibitem[{{Rasmussen} {et~al.}(2006){Rasmussen}, {Ponman}, {Mulchaey}, {Miles},
  \& {Raychaudhury}}]{Rasmussen2006}
{Rasmussen}, J., {Ponman}, T.~J., {Mulchaey}, J.~S., {Miles}, T.~A., \&
  {Raychaudhury}, S. 2006, \mnras, 373, 653,
  \dodoi{10.1111/j.1365-2966.2006.11023.x}

\bibitem[{{Richter} \& {Sancisi}(1994)}]{Richter94}
{Richter}, O.-G., \& {Sancisi}, R. 1994, \aap, 290, L9

\bibitem[{{Rupke} {et~al.}(2010){Rupke}, {Kewley}, \& {Chien}}]{Rupke10}
{Rupke}, D.~S.~N., {Kewley}, L.~J., \& {Chien}, L.-H. 2010, \apj, 723, 1255,
  \dodoi{10.1088/0004-637X/723/2/1255}

\bibitem[{{S{\'a}nchez} {et~al.}(2014){S{\'a}nchez}, {Rosales-Ortega},
  {Iglesias-P{\'a}ramo}, {Moll{\'a}}, {Barrera-Ballesteros}, {Marino},
  {P{\'e}rez}, {S{\'a}nchez-Blazquez}, {Gonz{\'a}lez Delgado}, {Cid Fernand
  es}, {de Lorenzo-C{\'a}ceres}, {Mendez-Abreu}, {Galbany}, {Falcon-Barroso},
  {Miralles-Caballero}, {Husemann}, {Garc{\'\i}a-Benito}, {Mast}, {Walcher},
  {Gil de Paz}, {Garc{\'\i}a-Lorenzo}, {Jungwiert}, {V{\'\i}lchez},
  {J{\'\i}lkov{\'a}}, {Lyubenova}, {Cortijo-Ferrero}, {D{\'\i}az}, {Wisotzki},
  {M{\'a}rquez}, {Bland-Hawthorn}, {Ellis}, {van de Ven}, {Jahnke},
  {Papaderos}, {Gomes}, {Mendoza}, \& {L{\'o}pez-S{\'a}nchez}}]{Sanchez14}
{S{\'a}nchez}, S.~F., {Rosales-Ortega}, F.~F., {Iglesias-P{\'a}ramo}, J.,
  {et~al.} 2014, \aap, 563, A49, \dodoi{10.1051/0004-6361/201322343}

\bibitem[{{Sancisi} {et~al.}(2008){Sancisi}, {Fraternali}, {Oosterloo}, \& {van
  der Hulst}}]{Sancisi08}
{Sancisi}, R., {Fraternali}, F., {Oosterloo}, T., \& {van der Hulst}, T. 2008,
  \aapr, 15, 189, \dodoi{10.1007/s00159-008-0010-0}

\bibitem[{{Schaefer} {et~al.}(2017){Schaefer}, {Croom}, {Allen}, {Brough},
  {Medling}, {Ho}, {Scott}, {Richards}, {Pracy}, {Gunawardhana}, {Norberg},
  {Alpaslan}, {Bauer}, {Bekki}, {Bland-Hawthorn}, {Bloom}, {Bryant}, {Couch},
  {Driver}, {Fogarty}, {Foster}, {Goldstein}, {Green}, {Hopkins},
  {Konstantopoulos}, {Lawrence}, {L{\'o}pez-S{\'a}nchez}, {Lorente}, {Owers},
  {Sharp}, {Sweet}, {Taylor}, {van de Sande}, {Walcher}, \&
  {Wong}}]{Schaefer2017}
{Schaefer}, A.~L., {Croom}, S.~M., {Allen}, J.~T., {et~al.} 2017, \mnras, 464,
  121, \dodoi{10.1093/mnras/stw2289}

\bibitem[{{Scott} {et~al.}(2018){Scott}, {Brinks}, {Cortese}, {Boselli}, \&
  {Bravo-Alfaro}}]{Scott2018}
{Scott}, T.~C., {Brinks}, E., {Cortese}, L., {Boselli}, A., \& {Bravo-Alfaro},
  H. 2018, \mnras, 475, 4648, \dodoi{10.1093/mnras/sty063}

\bibitem[{{Serra} {et~al.}(2012){Serra}, {Oosterloo}, {Morganti}, {Alatalo},
  {Blitz}, {Bois}, {Bournaud}, {Bureau}, {Cappellari}, {Crocker}, {Davies},
  {Davis}, {de Zeeuw}, {Duc}, {Emsellem}, {Khochfar}, {Krajnovi{\'c}},
  {Kuntschner}, {Lablanche}, {McDermid}, {Naab}, {Sarzi}, {Scott}, {Trager},
  {Weijmans}, \& {Young}}]{Serra12}
{Serra}, P., {Oosterloo}, T., {Morganti}, R., {et~al.} 2012, \mnras, 422, 1835,
  \dodoi{10.1111/j.1365-2966.2012.20219.x}

\bibitem[{{Spindler} {et~al.}(2018){Spindler}, {Wake}, {Belfiore}, {Bershady},
  {Bundy}, {Drory}, {Masters}, {Thomas}, {Westfall}, \& {Wild}}]{Spindler2018}
{Spindler}, A., {Wake}, D., {Belfiore}, F., {et~al.} 2018, \mnras, 476, 580,
  \dodoi{10.1093/mnras/sty247}

\bibitem[{{Springob} {et~al.}(2005){Springob}, {Haynes}, {Giovanelli}, \&
  {Kent}}]{Springob05}
{Springob}, C.~M., {Haynes}, M.~P., {Giovanelli}, R., \& {Kent}, B.~R. 2005,
  \apjs, 160, 149, \dodoi{10.1086/431550}

\bibitem[{{Stark} {et~al.}(2013){Stark}, {Kannappan}, {Wei}, {Baker}, {Leroy},
  {Eckert}, \& {Vogel}}]{Stark13}
{Stark}, D.~V., {Kannappan}, S.~J., {Wei}, L.~H., {et~al.} 2013, \apj, 769, 82,
  \dodoi{10.1088/0004-637X/769/1/82}

\bibitem[{{Stark} {et~al.}(2009){Stark}, {McGaugh}, \& {Swaters}}]{Stark2009}
{Stark}, D.~V., {McGaugh}, S.~S., \& {Swaters}, R.~A. 2009, \aj, 138, 392,
  \dodoi{10.1088/0004-6256/138/2/392}

\bibitem[{{Stark} {et~al.}(2021){Stark}, {Masters}, {Avila-Reese}, {Riffel},
  {Riffel}, {Boardman}, {Zheng}, {Weijmans}, {Dillon}, {Fielder}, {Finnegan},
  {Fofie}, {Goddy}, {Harrington}, {Pace}, {Rujopakarn}, {Samanso}, {Shamsi},
  {Sharma}, {Warrick}, {Witherspoon}, \& {Wolthuis}}]{Stark21}
{Stark}, D.~V., {Masters}, K.~L., {Avila-Reese}, V., {et~al.} 2021, \mnras,
  503, 1345, \dodoi{10.1093/mnras/stab566}

\bibitem[{{Tonnesen} {et~al.}(2023){Tonnesen}, {DeFelippis}, \&
  {Tuttle}}]{Tonnesen2023}
{Tonnesen}, S., {DeFelippis}, D., \& {Tuttle}, S. 2023, \apj, 951, 16,
  \dodoi{10.3847/1538-4357/acd3ee}

\bibitem[{{Torrey} {et~al.}(2012){Torrey}, {Cox}, {Kewley}, \&
  {Hernquist}}]{Torrey12}
{Torrey}, P., {Cox}, T.~J., {Kewley}, L., \& {Hernquist}, L. 2012, \apj, 746,
  108, \dodoi{10.1088/0004-637X/746/1/108}

\bibitem[{{Trapp} {et~al.}(2022){Trapp}, {Kere{\v{s}}}, {Chan}, {Escala},
  {Hummels}, {Hopkins}, {Faucher-Gigu{\`e}re}, {Murray}, {Quataert}, \&
  {Wetzel}}]{Trapp2022}
{Trapp}, C.~W., {Kere{\v{s}}}, D., {Chan}, T.~K., {et~al.} 2022, \mnras, 509,
  4149, \dodoi{10.1093/mnras/stab3251}

\bibitem[{{Tuttle} \& {Tonnesen}(2020)}]{Tuttle20}
{Tuttle}, S.~E., \& {Tonnesen}, S. 2020, \apj, 889, 188,
  \dodoi{10.3847/1538-4357/ab5dbb}

\bibitem[{{van de Voort} {et~al.}(2015){van de Voort}, {Davis}, {Kere{\v s}},
  {Quataert}, {Faucher-Gigu{\`e}re}, \& {Hopkins}}]{VanDeVoort15}
{van de Voort}, F., {Davis}, T.~A., {Kere{\v s}}, D., {et~al.} 2015, \mnras,
  451, 3269, \dodoi{10.1093/mnras/stv1217}

\bibitem[{{van den Bosch}(1998)}]{VandenBosch1998}
{van den Bosch}, F.~C. 1998, \apj, 507, 601, \dodoi{10.1086/306354}

\bibitem[{{van Eymeren} {et~al.}(2011){van Eymeren}, {J{\"u}tte}, {Jog},
  {Stein}, \& {Dettmar}}]{vanEymeren2011}
{van Eymeren}, J., {J{\"u}tte}, E., {Jog}, C.~J., {Stein}, Y., \& {Dettmar},
  R.~J. 2011, \aap, 530, A30, \dodoi{10.1051/0004-6361/201016178}

\bibitem[{{Watts} {et~al.}(2020){Watts}, {Catinella}, {Cortese}, \&
  {Power}}]{Watts2020}
{Watts}, A.~B., {Catinella}, B., {Cortese}, L., \& {Power}, C. 2020, \mnras,
  492, 3672, \dodoi{10.1093/mnras/staa094}

\bibitem[{{Watts} {et~al.}(2023){Watts}, {Cortese}, {Catinella}, {Power},
  {Fraser-McKelvie}, {Bryant}, {Croom}, {van de Sande}, {Bland-Hawthorn}, \&
  {Groves}}]{Watts2023}
{Watts}, A.~B., {Cortese}, L., {Catinella}, B., {et~al.} 2023, \mnras, 519,
  1452, \dodoi{10.1093/mnras/stac3643}

\bibitem[{{Welikala} {et~al.}(2008){Welikala}, {Connolly}, {Hopkins},
  {Scranton}, \& {Conti}}]{Welikala2008}
{Welikala}, N., {Connolly}, A.~J., {Hopkins}, A.~M., {Scranton}, R., \&
  {Conti}, A. 2008, \apj, 677, 970, \dodoi{10.1086/527666}

\bibitem[{{Westfall} {et~al.}(2019){Westfall}, {Cappellari}, {Bershady},
  {Bundy}, {Belfiore}, {Ji}, {Law}, {Schaefer}, {Shetty}, {Tremonti}, {Yan},
  {Andrews}, {Brownstein}, {Cherinka}, {Coccato}, {Drory}, {Maraston},
  {Parikh}, {S{\'a}nchez-Gallego}, {Thomas}, {Weijmans}, {Barrera-Ballesteros},
  {Du}, {Goddard}, {Li}, {Masters}, {Ibarra Medel}, {S{\'a}nchez}, {Yang},
  {Zheng}, \& {Zhou}}]{Westfall19}
{Westfall}, K.~B., {Cappellari}, M., {Bershady}, M.~A., {et~al.} 2019, \aj,
  158, 231, \dodoi{10.3847/1538-3881/ab44a2}

\bibitem[{{White} \& {Frenk}(1991)}]{White91}
{White}, S.~D.~M., \& {Frenk}, C.~S. 1991, \apj, 379, 52,
  \dodoi{10.1086/170483}

\bibitem[{{York} {et~al.}(2000){York}, {Adelman}, {Anderson}, {Anderson},
  {Annis}, {Bahcall}, {Bakken}, {Barkhouser}, {Bastian}, {Berman}, {Boroski},
  {Bracker}, {Briegel}, {Briggs}, {Brinkmann}, {Brunner}, {Burles}, {Carey},
  {Carr}, {Castander}, {Chen}, {Colestock}, {Connolly}, {Crocker}, {Csabai},
  {Czarapata}, {Davis}, {Doi}, {Dombeck}, {Eisenstein}, {Ellman}, {Elms},
  {Evans}, {Fan}, {Federwitz}, {Fiscelli}, {Friedman}, {Frieman}, {Fukugita},
  {Gillespie}, {Gunn}, {Gurbani}, {de Haas}, {Haldeman}, {Harris}, {Hayes},
  {Heckman}, {Hennessy}, {Hindsley}, {Holm}, {Holmgren}, {Huang}, {Hull},
  {Husby}, {Ichikawa}, {Ichikawa}, {Ivezi{\'c}}, {Kent}, {Kim}, {Kinney},
  {Klaene}, {Kleinman}, {Kleinman}, {Knapp}, {Korienek}, {Kron}, {Kunszt},
  {Lamb}, {Lee}, {Leger}, {Limmongkol}, {Lindenmeyer}, {Long}, {Loomis},
  {Loveday}, {Lucinio}, {Lupton}, {MacKinnon}, {Mannery}, {Mantsch}, {Margon},
  {McGehee}, {McKay}, {Meiksin}, {Merelli}, {Monet}, {Munn}, {Narayanan},
  {Nash}, {Neilsen}, {Neswold}, {Newberg}, {Nichol}, {Nicinski}, {Nonino},
  {Okada}, {Okamura}, {Ostriker}, {Owen}, {Pauls}, {Peoples}, {Peterson},
  {Petravick}, {Pier}, {Pope}, {Pordes}, {Prosapio}, {Rechenmacher}, {Quinn},
  {Richards}, {Richmond}, {Rivetta}, {Rockosi}, {Ruthmansdorfer}, {Sandford},
  {Schlegel}, {Schneider}, {Sekiguchi}, {Sergey}, {Shimasaku}, {Siegmund},
  {Smee}, {Smith}, {Snedden}, {Stone}, {Stoughton}, {Strauss}, {Stubbs},
  {SubbaRao}, {Szalay}, {Szapudi}, {Szokoly}, {Thakar}, {Tremonti}, {Tucker},
  {Uomoto}, {Vanden Berk}, {Vogeley}, {Waddell}, {Wang}, {Watanabe},
  {Weinberg}, {Yanny}, {Yasuda}, \& {SDSS Collaboration}}]{York00}
{York}, D.~G., {Adelman}, J., {Anderson}, Jr., J.~E., {et~al.} 2000, \aj, 120,
  1579, \dodoi{10.1086/301513}

\end{thebibliography}
\bibliographystyle{aasjournal}



\appendix 
\section{Flux Calibration}
\begin{figure}
    \plotone{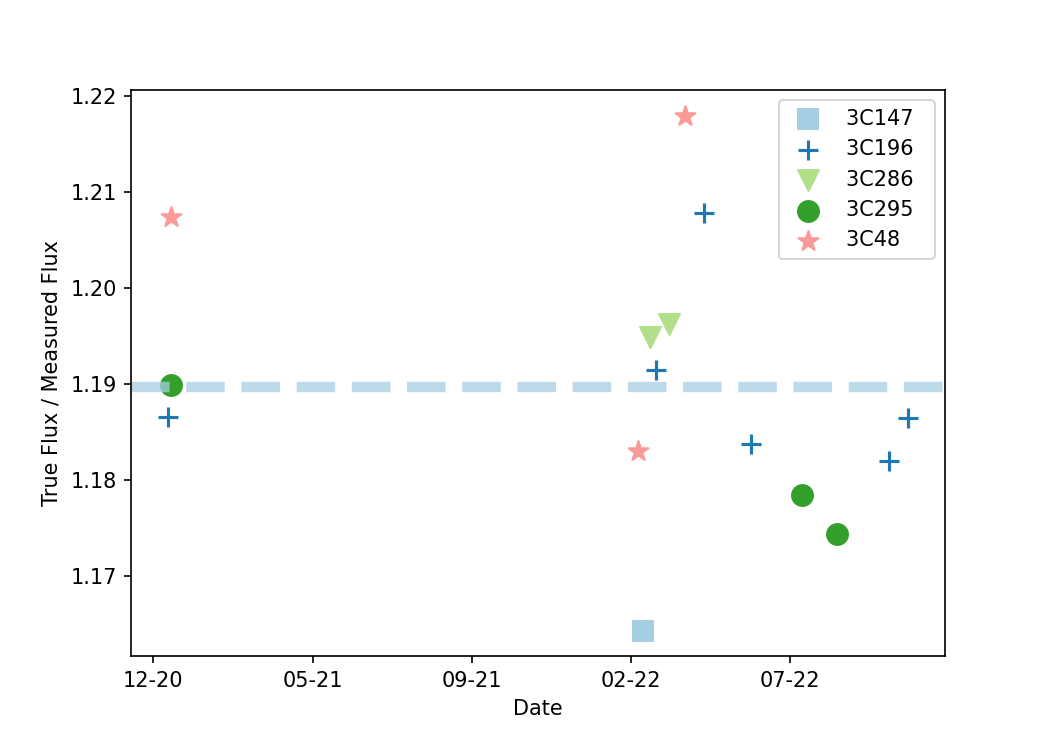}
    \caption{True source flux density at 1.4 GHz divided by measured source flux density using the default \texttt{GBTIDL} flux calibration as a function of time for observations of standard radio sources during programs \texttt{GBT20B-261} and \texttt{GBT22A-436}. The horizontal dashed line shows the mean of the data.}
    \label{fig:calibration}
\end{figure}

On a cadence on every four days (or the spacing between observing sessions, if they were separated by more than four days), we carried out spectroscopic observations of bright radio sources in order to check the telescope pointing solution and calibrate the flux scale. These observations were done using position switching with scan times of 30 seconds. The data from both polarizations were averaged and the mean flux measured over the bandpass while avoiding the central region of the spectrum dominated by Milky Way emission. We applied that default \texttt{GBTIDL} flux calibration and the resulting flux densities were compared to the true flux densities of measured calibrators taken from \citet{Ott1994}, or \citet{Perley2017} if available. The results are shown in Figure~\ref{fig:calibration}. A subset of targets that we observed (Virgo A, Hydra A, 3C227, 3C348, 3C353) are not shown in Figure~\ref{fig:calibration} due to them being potentially poor flux calibrators (see \citealt{Ott1994} and \citealt{Perley2017} for reasoning). 

We find the ratio of true flux density to measured flux density (using the default \texttt{GBTIDL} calibration coefficients) is $1.190+/-0.003$. Our results are highly consistent with a recent study by \citet{Goddy20} who conducted a similar analysis using data from 2016 to 2019 and found an offset of 20\% in the true-to-measured flux density of bright calibrators. Based on these results, we multiply all our measured fluxes and flux densities by 1.2 after processing our data using the standard GBTIDL calibration.

\end{document}